\documentclass[12pt]{article}
\usepackage{amsfonts,epsfig,subfigure,amsxtra,amssymb,amsmath,graphics}
\usepackage{epstopdf}
\usepackage{multirow}
\usepackage{indentfirst,bm,mathrsfs}
\usepackage{color}



\numberwithin{equation}{section}

\def\qed{\hfill{\vrule height 1.9ex width .9ex depth -.1ex }}

\newcounter{intheorem}

\def\({\left(}
\def\){\right)}

\DeclareGraphicsExtensions{.pdf} 

\begin{document}

\title{{\bf \ \ \  The FFBS Estimation of High-dimensional Panel Data Factor Stochastic Volatility Model}}
\author{ Guobin Fang \thanks{ School of Statistics and Applied Mathematics,
 Anhui University of Finance and Economics, Bengbu 233030, Anhui,  China
 }
\and
Huimin Ma \thanks{ School of Statistics and Applied Mathematics,
 Anhui University of Finance and Economics, Bengbu 233030, Anhui,  China}
\and
Michelle Xia \thanks{ Division of Statistics, Northern Illinois University, Dekalb 60115, IL, USA}
\and
Bo Zhang \thanks{ Center for Applied Statistics, Institute of Probability and Statistics, School of Statistics,
 Renmin University of China, Beijing 100872, China}
}

\date{}

\maketitle

\thispagestyle{empty}

\begin{abstract}

In this paper, we propose a novel panel data factor stochastic volatility model embracing both observable factors and unobservable factors that can influence financial returns. The model is an extension of the multivariate stochastic volatility model composed of the three parts of the mean equation, the volatility equation and the factor volatility evolution. The stochastic volatility equation uses a one-step forward prediction process that simplifies parameter estimation in high-dimensional parameter spaces. In particular, Forward Filtering Backward Sampling (FFBS) based on the Kalman Filter Recursive Algorithm (KFRA) is utilized for parameter estimation of the stochastic volatility equation, under the Bayesian Markov Chain Monte Carlo Simulation (MCMC) framework. The results from numeric simulations demonstrate that the algorithm possesses robustness and consistency for parameter estimation and latent factor sampling. Using the new model and its estimation method, this paper makes a comparative analysis on the influence of observable and unobservable factors on Internet financial and traditional financial companies listed in the Chinese stock market. The results show that the influence of the observable factors is similar for the two types of listed companies, while that of the unobservable factors is noticeably different.

\end{abstract}

\newpage

\newpage
\setlength{\baselineskip}{0.28in}

\section{\sc {Introduction}}

In the financial markets, the return and volatility of financial assets are both sequentially correlated, with the volatility typically having time-varying characteristics. Both the stochastic volatility (SV) model and generalized autoregressive conditional heteroskedasticity (GARCH) model have been widely used to capture the dependence and the volatility clustering in financial time series. The stochastic volatility (SV) model assumes that the stochastic volatility of the return is influenced by unobservable factors, accompany with the impact from the volatility of the previous period. The stochastic volatility model is shown to perform well in modeling the realized volatility of high frequency data. From the in-depth study of the realized volatility, high frequency transaction data constantly reflect the continuous time characteristics, giving rise to the majority of the modern stochastic volatility research carried out around continuous data. In particular, the geometric Brownian motion process, the non-Gaussian O-U process, the time-varying Levy process, the Markovian switching process and other stochastic processes with jump based on continuous time have been extensively applied for stochastic volatility models.

Shephard (2005) marked the beginning of a new generation of stochastic volatility models. Whereas the early stochastic volatility models were simple stochastic differential equations based on the geometric Brownian motion, the new generation of research mainly focused on update methods for model estimation and testing. Since the late 1990s, there have been great advances in stochastic volatility models for the purposes of realized volatility prediction and high frequency modeling. In addition to traditional estimation and testing, recent research on stochastic volatility models focused more on the characterization of local characteristics, such as long memory, jump, diffusion and microstructure noise, in the modeling of high frequency transaction data. These stochastic volatility models can process both continuous and discrete data.

Although there is some controversy about who firstly proposed the stochastic volatility model (see Shephard (2005) for details), it is an indisputable fact that the stochastic volatility model have been applied to discrete time series analysis. Due to the incomparability of high frequency and relatively low frequency data in terms of means of acquisition and analysis significance, the study of univariate stochastic volatility model originated from the application of time-varying Brownian motion in financial econometrics. Taylor (1982) is regarded as one of the early pioneers of stochastic volatility model for his contributions in introducing the features of volatility clustering and time-varying volatility simultaneously. Harvey et al. (1994) first proposed the multivariate stochastic volatility model. While Diebold and Nerlove's (1989) adopted multiple factors in their multivariate factor model, the multivariate stochastic volatility model lays more emphasis on the study of multiple financial assets. In addition to the asymmetry caused by the difference in reaction to risk, the multivariate  random volatility model also considers the joint influence of multiple assets. Therefore, the multivariate factor model plays an important role in multivariate stochastic volatility modeling.

Han (2005) and Chib et al. (2006) pushed the research of high-dimensional multivariate stochastic volatility modeling further. In their studies, the high-dimensional feature contains two aspects. On the one hand, the high dimension refers to the large number of assets, including dozens or even hundreds of financial assets. On the other hand, there are a large number of parameters that need to be estimated due to the complexity of the multivariate stochastic volatility model structure. In the Han's study, a total of 222 parameters need to be estimated in the stochastic volatility model when there are 36 return series and 3 factors. In the Chib's model, if the number of assets is 50 and the number of factors is 8, then the parameters to be estimated will reach 688. The estimation of such a large number of parameters requires a special computation method. The high-dimensional factor stochastic volatility models all adopted the block estimation method. The block estimation method reduces the complexity of calculation by dividing the parameters into several groups for block updating in the Gibbs sampler or the Metropolis-Hastings algorithm.

For the stochastic volatility model, the covariance structure of the model's error components must be specified. For the error components, routine rules regarding positive definitiveness and correlation must be followed regarding the covariance matrix. Tsay (2010) considered Cholesky decomposition to re-parametrize the covariance matrix owing to its many advantages. Firstly, Cholesky decomposition ensures the positive definitiveness of a covariance matrix easily. Secondly, each element of the decomposition matrix has favorable explanatory ability. Finally, such decomposition can reflect the time-varying characteristics of volatility. Since Cholesky decomposition can run into high-dimensional problems, Lopes, McCulloch and Tsay (2012) put forward a more practical approach to estimate the Cholesky's stochastic volatility model. Their method adopts the matrix transformation of the multivariate random error terms, and uses Cholesky decomposition to simplify the estimation of the large number of parameters for the conditional covariance matrix. Although these methods simplify the structure of random error components, corresponding estimation methods are required for processing the high-dimensional parameters. For this purpose, Kastner et al. (2017) proposed an efficient Bayesian method to estimate the factor stochastic volatility model, which they demonstrated the superior performance on the analysis of 26-dimensional exchange rate data.

In contrast to the univariate stochastic volatility model that studies the influencing factors for the rate of return on a single asset, the multivariate stochastic volatility model also captures the influences on a large number of financial assets. For both the univariate and multivariate stochastic volatility models, these influencing factors are considered to be either observable or unobservable. The influences from these factors are considered as self-driven. In fact, in addition to the influencing factors on the characteristics of volatility, the impact on the asset price changes can also be analyzed for external influencing factors. For example, some models solely consider the influence of observable factors on the change of asset return, which include the Fama and French's three-factor model (1992) and the BARRA factor model (proposed by BARRA company).

In order to study the influence of observable factors and unobservable factors on the stochastic volatility of financial assets furtherly, this paper proposes the panel data stochastic volatility model by introducing observable factors into the multivariate stochastic volatility model. Such observable factors can include not only market factors such as industrial (stock) average return, trading volume and transaction amount, but also out of counter trade factors such as macroeconomic variables and industrial development level. Compared with the regular multivariate stochastic volatility model, the panel data stochastic volatility model considers a large number of observable factors in the market and out of the market, providing a more realistic practical application. Despite the theoretical basis and practical application significance inherited from the stochastic volatility model, add-on observable influencing factors can lead to more difficult in the model estimation and interpretation in the panel data stochastic volatility model.

This paper proposes a novel panel data factor stochastic volatility model and estimation methods for the high-dimensional parameters. The main contents are arranged as follows. The second section discusses the specification of the panel data factor stochastic volatility model, including the mean equation, volatility equation and factor equation, along with the embodiment of the stochastic effect. The third section studies model estimation methods, including the realization of joint estimation based on MCMC and FFBS algorithm. The fourth section presents simulation studies to test the validity of the proposed estimation method. The fifth section performs an application study for comparing internet financial and traditional financial companies listed in China's stock market, based on the differences in the observable and unobservable factors that influence the two types of companies. Finally, concluding remarks and future research are given in the last section.

\section{\sc {Specifications of Panel Data Factor Stochastic Volatility Model}}

The specification of stochastic volatility term in panel data stochastic volatility model (PDSVM) is similar to the multivariate stochastic volatility model. The factors related to the change of the rate of return are treated as explanatory variables in the model, which will influence the structure of the stochastic volatility terms. Since the multivariate factor model is difficult to explain theoretically and cannot adapt to model transformation, we only consider the additive factor structure for the panel data factor stochastic volatility model. Furthermore, the additive factors and other covariates are assumed to be independent each other. For the individual and period effects, we propose to use either the panel data random effect model where such effects are treated as random effects, or the panel data fixed effect model where they are treated as fixed effects. In both cases, the individual effect and the time effect can be included simultaneously in the model.

\subsection{Panel Data Stochastic Volatility Model}

In financial asset allocation and portfolio management, the excess logarithmic return rate of an investment portfolio with $N$ financial assets can be denoted by ${\bf{r}}_t  =$ $(r_{1t} ,r_{2t} , \cdots ,r_{Nt} )'$ , where $r_{it}$ $(i = 1,2, \cdots ,N, t = 1, \cdots ,T)$ represents the excess logarithmic return rate of asset $i$ in time $t$. The varying coefficient panel data random effect model can be used to capture the influence of observable and unobservable factors on the return rate of financial assets. The mean equation of panel data stochastic volatility model can be set as follows:
\begin{equation}\label{eq2.1}
 r_{it}  = {\bm{\beta }}_i^{'} {\bf{x}}_{it}  + \xi _i  + \eta _t  + u_{it}, 
\end{equation}
where the vector ${\bf{x}}_{it}$ contains the observable factors that affect the rate of return on financial assets, with its dimension $k$ being the number of influence factors that can be either internal or external factors affecting the financial market. The vectors $\xi _i$ and $\eta _t$ contain respectively the individual effects and the time effects that are both assumed to be random. The influence of the factors on the volatility of $r_{it}$ is reflected in the specification of error component terms. Here, we focus on the individual effects. Assuming that the mean of the error component $u_{it}$ at time $t$ is  $0$, ${\bm{\mu }}_\varepsilon   = E({\bf{u}}_t ) = {\bf{0}}$, the conditional covariance matrix satisfies $Cov({\bf{u}}_t )={\bm{\Sigma }}_t  = diag(\sigma _{1t}^2 , \cdots ,\sigma _{Nt}^2 )$ with $\sigma _{it}^2  = \exp (h_{it} )$.

Furthermore, we assume that the volatility equation of the panel data stochastic volatility model is
\begin{equation}\label{eq2.2}
h_{it}  = \alpha _{i0}  + \alpha _{i1} h_{it - 1}  + v_{it},
\end{equation}

where $\alpha _{i0}$ is a scalar, $u_{it}$ and $v_{it}$ have independent normal distributions with mean $0$ and variance $\sigma _{i\xi }^2$ and $\sigma _{i\eta }^2$ respectively. The exponential transformation of $h_{it}$ guarantees the positive definiteness of the variance covariance matrix ${\bf{\Sigma }}_t$. In order to satisfy the stationarity of the time series, it is assumed that the regression parameter satisfies $\left| {\alpha _{i1} } \right| < 1$, otherwise, alternatively higher-order lag terms of the corresponding variable can be introduced. Models (\ref{eq2.1}) and (\ref{eq2.2}) constitute the basic form of a panel data stochastic volatility model. The random error term is simplified via specification of the random individual effect and random time effect in the panel data model.

\subsection{Panel Data Factor Stochastic Volatility Model}

The explanatory variables in panel data stochastic volatility model (\ref{eq2.1}) only reflect observable factors, which may be internal or external market factors. In the aforementioned model, these influencing factors constitute the explanatory variables of the panel data model. The unobservable factors of the panel data model comprise of three parts: the random effect (fixed effect), random error component (stochastic volatility term), and statistical factor component. Here, panel data factor stochastic volatility model (PDFSVM) utilities the statistical factor stochastic volatility model to capture the influence of common shock on multiple assets. This common shock is represented by common factors. After introducing the common factors, the panel data factor stochastic volatility model can be written as
\begin{equation}\label{eq2.3}
r_{it}  = {\bm{\beta }}_i^{'} {\bf{x}}_{it}  + {\bm{\lambda }}_i^{'} {\bf{f}}_t  + u_{it},
\end{equation}
where the random error component possesses the structure in (\ref{eq2.2}). In the factor decomposition term ${\bm{\lambda }}_i^{'} {\bf{f}}_t$, ${\bm{\lambda }}_i  = (\lambda _{i1} , \cdots ,\lambda _{ip} )^{'}$ contains the factor loadings, and ${\bf{f}}_t  = (f_{1t} , \cdots ,f_{pt} )^{'}$ represents the common factors, with the number of common factors is $p$ ($p < N$).

The conditional covariance structure of stochastic volatility term $u_{it}$ is the same as the panel data stochastic volatility model (\ref{eq2.1}). In order to reflect the lag effect of common factors, i.e., the impact of continuous decay caused by the common shocks, we specify a structure similar to the multi-factor stochastic volatility model proposed by Jacquier et al. (1995) and Lopes and Carvalho (2007).

In particular, we assume that the common factors of panel data stochastic volatility model have the same evolution of stochastic volatility:
\begin{equation}\label{eq2.4}
\begin{aligned}
f_{jt}  = \exp (q_{jt} /2)\varepsilon _{jt}\\
q_{jt}  = \varphi _{j0}  + \varphi _{j1} q_{jt - 1}  + w_{jt}\\
j = 1, \cdots ,p;\quad t = 1, \cdots ,T,
\end{aligned}
\end{equation}
where $\varepsilon _{jt}  \sim N(0,1)$, $w_{jt}  \sim N(0,\sigma _w^2 )$, and the error terms are independent each other. Based on the process (\ref{eq2.4}), for given $q_{jt}$ and any $j = 1, \cdots ,p$, we have $E(f_{jt} ) = 0$ and $Var(f_{jt} ) = \exp (q_{jt} )$. The common factors in ${\bf{f}}_t$ satisfy $({\bf{f}}_t |{\bf{Q}}_t ) \sim N({\bf{0}},{\bf{Q}}_t )$, where ${\bf{Q}}_t$  is the variance-covariance matrix of the common factors ${\bf{f}}_t$. Hence, we have ${\bf{Q}}_t  = diag(\exp (q_{1t} ), \cdots ,\exp (q_{pt} ))$ for the conditional variance, where $q_{jt}$ is given by the AR (1) process in equation (\ref{eq2.4}). In order to ensure the stationarity of the sequence generated by the state process (\ref{eq2.4}), it is assumed $|\varphi _{j1} | < 1$ here.

Equations (\ref{eq2.3}), (\ref{eq2.2}) and (\ref{eq2.4}) compose the panel data factor stochastic volatility model. Due to some restrictions applied to the factor loadings and common factors in factor decomposition, the estimation of panel data factor stochastic volatility model is more complex than the regular panel data stochastic volatility model. The estimation of the effects of the predictive variables and latent variables in model (\ref{eq2.2}) - (\ref{eq2.4}) need to be carried out simultaneously, a situation difficult for obtaining a closed-form solution for the maximum likelihood estimation. For numerical optimization based on iterative algorithms, the large number of parameters gives rise to difficulty in the algorithm convergence in the high-dimensional parameter space.

In the panel data factor stochastic volatility model, the curse of dimensionality mainly comes from the large number of parameters to be estimated for the conditional covariance matrix and factor decomposition. Although the application of the factor model can help to reduce the dimensionality when the number of assets $N$ is very large, the total number of parameters generated from the estimation process can be far greater than the number of assets. For the panel data factor stochastic volatility model (\ref{eq2.3}), ${\bm{\beta }}_i$ is a $k$ dimensional vector, up to a total of $k \times N$ regression coefficients. ${\bm{\lambda }}_i$ and ${\bf{f}}_t$ are $p$ dimensional column vectors. Since the number of periods and individuals are $T$ and $N$ respectively, under the condition of applying identification constraints, ${\bm{\lambda }}_i$ and ${\bf{f}}_t$ have $p(p - 1)/2$ and $p(p + 1)/2$ constraints respectively. With ${\bf{f}}_t$ being latent factors, and there are still $Np - (p^2  + p)/2$ free parameters to be estimated. The number of coefficients in model (\ref{eq2.4}) $\alpha _{i0}$ and $\alpha _{i1}$ is equal to the total number of financial assets $N$. Both $\varphi _{j0}$ and $\varphi _{j1}$ in model (\ref{eq2.4}) contain coefficients for the regression of AR (1), each of them having $p$ parameters to be estimated.

For applying identification constraints, there are $kN + Np - (p^2  + p)/2 + 2(N + p)$ parameters to be estimated in the panel data factor stochastic volatility model (\ref{eq2.2}) - (\ref{eq2.4}), $Np - (p^2  + p)/2$ of which are determined by factor decomposition. Thus, for a financial portfolio with $N = 40$ assets, $k=3$ explanatory variables, and $p=6$ factors, there are 431 model parameters to be estimated
without consideration the factor loading coefficients and the parameters related to random error terms. For the panel data factor stochastic volatility model, the high-dimensional problem can be caused by a large value of either $k$,  $N$, $T$, or $p$. Hence, we will consider how to reasonably reduce the dimension for model computation and parameter estimation in high-dimensional scenarios.

From the above specification, we may argue that the panel data stochastic volatility model is similar to the panel data factor stochastic volatility model, despite their difference in the factor model structure. In the subsequent sections, we will focus on the panel data factor stochastic volatility model, although many of the results are also applicable to the regular panel data stochastic volatility model. From a comparative standpoint, the model specification and estimation is more complicated for the panel data factor stochastic volatility model in high-dimensional scenarios, and which will require innovative iterative algorithms and computing techniques.

\section{\sc {Estimation and Computation of Panel Data Factor Stochastic Volatility Model}}

Model (\ref{eq2.3}) is an extension of the panel data dynamic mixed double factor model (DMDFM) proposed by Fang, Zhang and Chen (2018). For the mean equation of the model (\ref{eq2.3}), Bai (2009) developed the Least Squares (LS) method and the Least Squares Dummy Variable (LSDV) estimator of panel data fixed effect model based on Pesaran (2006) and Coakley, et al. (2002), he proved the consistency, asymptotic normality and other limiting properties of estimators. Similar alternative estimation methods include the Generalized Moment Method (GMM) of Ahn, Lee and Schmidt (2001) and the Quasi-Difference method (QD) of Holtz-Eakin, Newey and Rosen (1988). For the time series dynamic factor model that contains a large number of predictors, Stock and Watson (2002) proposed Nonlinear Least Square method to estimate the model parameters.

The parameter estimation of the panel data factor stochastic volatility model becomes relatively complex after adding the stochastic equation (\ref{eq2.2}) and (\ref{eq2.4}). In particular, the likelihood function associated with the model is intractable. Since models (\ref{eq2.2}), (\ref{eq2.3}) and (\ref{eq2.4}) are nested with each other, it is natural to perform estimation based on the profile likelihood given the estimates of the nested parameters. This paper uses MCMC method based on Gibbs sampling and Metropolis-Hastings Algorithm that allows block updating in posterior sampling.

\subsection{Preliminary Tact}

The observable information of the panel data factor stochastic volatility model can be divided into two parts. The first part comes from the predictors (independent variables) ${\bf{x}}_{it}$, and the other part comes from the response (dependent) variable $r_{it}$. Denoting the information set  $\mathscr{I} _{t - 1}$ composed of observable historical records, the parameters of the model are $\boldsymbol{\omega}  = ({\bf{\beta }}_i ,{\bf{\lambda }}_i ,{\bf{f}}_t ,\alpha _{i0} ,\alpha _{i1} ,\varphi _{j0} ,\varphi _{j1} )$, and the conditional density function of the latent variables is $p(\boldsymbol{\psi} |\boldsymbol{\omega} ,\mathscr{I}_{t - 1})$. The conditional probability function of the observed variables (i.e., the conditional likelihood function with respect to the parameters) can be written as:
\begin{equation}\label{eq5}
\begin{aligned}
p({\bf{r}}_t |\boldsymbol{\omega} ) & = \mathop \prod \nolimits_{t = 1}^T \iint p({\bf{r}}_{\bf{t}} |{\bf{h}}_t ,{\bf{q}}_t ,{\bm{\beta }}_i ,{\bm{\lambda }}_i ,{\bf{f}}_t ,{\bf{x}}_t )p(\boldsymbol{\psi} |\boldsymbol{\omega} ,\mathscr{I}_{t - 1} )d{\bf{h}}_t d{\bf{q}}_t\\
& = \mathop \prod \nolimits_{t = 1}^T \iint N({\bf{r}}_{\bf{t}} |{\bm{\beta }}_i^{'} {\bf{x}}_{it}  + {\bm{\lambda }}_i^{'} {\bf{f}}_t ,{\bm{\Omega }}_t )p(\boldsymbol{\psi} |\boldsymbol{\omega} ,\mathscr{I}_{t - 1} )d{\bf{h}}_t d{\bf{q}}_t,
\end{aligned}
\end{equation}
where, $N( \cdot | \cdot )$ denotes the multivariate normal distribution, ${\bm{\beta }}_i^{'} {\bf{x}}_{it}  + {\bm{\lambda }}_i^{'} {\bf{f}}_t$ is the conditional mean of ${\bf{r}}_{\bf{t}}$, and ${\bm{\Omega }}_t$ is its marginal condition covariance matrix that can be written as:
\begin{equation}\label{eq6}
{\bm{\Omega }}_t  = {\bm{\Lambda Q}}_t {\bm{\Lambda}}^{'} + {\bm{\Sigma }}_t.
\end{equation}
Although the density function of the multivariate normal distribution has an explicit form, the above integral in (\ref{eq5}) does not have an analytical solution. For the proposed model, the conditional density function of latent variables $\boldsymbol{\psi}$ can't be written analytically. Thus, the likelihood function given in equation (\ref{eq5}) in tractable, making it difficult to estimate the panel data factors stochastic volatility model via maximum likelihood estimation.

Here, we will use the Markov Chain Monte Carlo Simulation (MCMC) method to implement the panel data factor stochastic volatility model (\ref{eq2.1}) - (\ref{eq2.4}). This method constructs aperiodic and irreducible Markov chains to obtain the invariant-distribution of posterior distribution of the target parameters. For the proposed model, the latent variables and parameters can be sampled simultaneously based on Markov chains. In particular, the joint posterior distribution can be denoted as follows:
\begin{equation}\label{eq7}
\pi ({\bm{\beta }}_i ,{\bm{\lambda }}_i ,{\bf{f}}_t ,\alpha _{i0} ,\alpha _{i1} ,\varphi _{j0} ,\varphi _{j1} ,h_{it} ,q_{jt} |{\bf{r}}_t ,{\bf{x}}_t ).
\end{equation}
The invariant-distribution of all latent variables and parameters will contain a huge number of distributional components that can be decomposed for computational purposes. For the Bayesian implementation, we will resort to hierarchical Bayesian methods for which the effectiveness and consistency of estimation depends on the prior specification and other posterior settings. Because of the dynamic features of the parameters and the similarity in the structure of the AR (1) process in (\ref{eq2.2}) and (\ref{eq2.4}), the dynamic correlation of the latent variables will be estimated via the processing method of the Dynamic Linear Model (DLM).

With adaption to the high-dimensional characteristics of the multivariate stochastic volatility model, Chib (2001) summarized the Bayesian inference method based on MCMC techniques. Chib, Nardari and Shephard (2006) discussed the estimation and comparison of high-dimensional latent factor stochastic volatility models with jumps and alternative specifications. Han (2005) studied the portfolio construction and risk control of a large number of financial assets by using the dynamic factor multivariate stochastic volatility model, and achieved favorable prediction performance. For the high-dimensional multivariate stochastic volatility model and the panel data stochastic volatility model, estimation of a large number of parameters in the presence of latent variables is a challenging problem in Bayesian inference. Lopes, McCulloch and Tsay (2012) used parallel computing techniques. In order to obtain computational efficiency, they performed parallel estimation of multiple assets via recursive conditional regression, after partition the assets into several smaller portfolios. The parallel computing technique is appropriate for the multivariate factor decomposition model. For the panel data stochastic volatility model, however, the explanatory variables exhibit both temporal and spatial correlation, resulting in cross-sectional dependence among individuals that needs to be captured using alternative algorithms.

In the Bayesian estimation of the panel data stochastic volatility model, Chib et al. (2006) proposed the blocking method to improve the computing speed. The method divides the parameters and latent variables into separate blocks (groups) so that they can be updated sequentially within each iteration of the MCMC algorithm. The blocking algorithm design requires consideration of factors such as the blocking strategy, the form of the likelihood function, the prior distribution specification, and the form of the joint posterior distribution of the parameters.

For designing a blocking MCMC algorithm, it is natural to divide the parameters and latent variables in (\ref{eq5}) into three parts according to the three equations of the panel data factor stochastic volatility model. The three parts are the parameters and variables $({\bm{\beta }}_i ,{\bm{\lambda }}_i ,{\bf{f}}_t ,h_{it} ,q_{jt} )$ in model (\ref{eq2.3}); $(\alpha _{i0} ,\alpha _{i1} )$, the mean and variance of the random terms in model (\ref{eq2.2}); $(\varphi _{j0} ,\varphi _{j1} )$ in model (\ref{eq2.4}) and the estimation of mean and variance of its random term. When subdivided further, $({\bm{\beta }}_i )$, $({\bm{\lambda }}_i )$, $({\bf{f}}_t ,q_{jt} )$ and $(h_{it} )$ can be sampled in blocks. For posterior sampling, the conditional posterior distribution of each parameter or latent variable must be considered. When using the hierarchical Bayesian method, both the specific form and parameters of the distribution needs to be studied. In the subsequent paragraphs, we will focus on the algorithm implementation, after specifying the posterior parameters. In order to reduce computational time, each iteration of the MCMC algorithm will be divided into three blocks corresponding to the estimation of regression coefficients, factor decomposition and the dynamic linear model. The relationship between these three parts is shown in the simplified directed graphical structure given in Figure 1.

\begin{figure}[ht]
\includegraphics[bb=15 250 0 500,scale=.85]{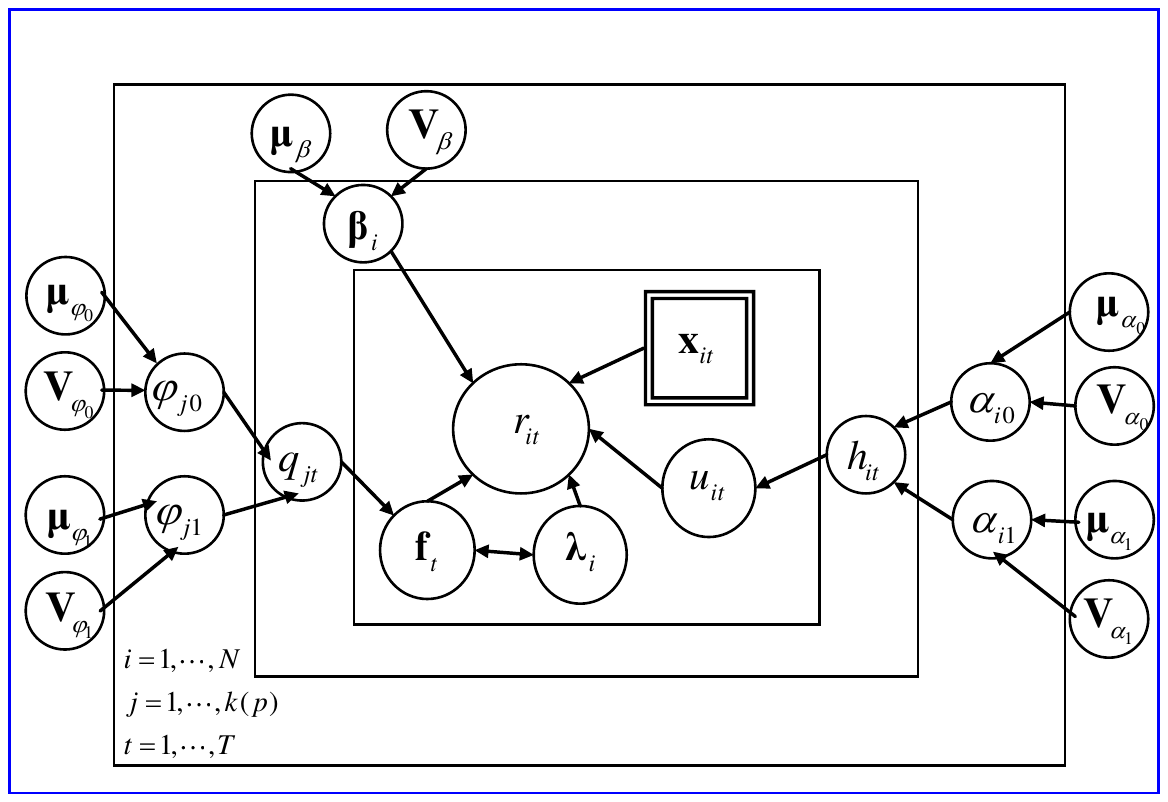}
\caption{Directed graphical structure for the panel data factor stochastic volatility model.}
\end{figure}

\subsection{Prior and Posterior Specification for Latent Variables and Model Parameters }

The key step in Bayesian block sampling techniques is the specification of the ''blocks'', and the conditional posterior distribution of each block whilst the others are given. For the panel data factor stochastic volatility model, the parameters and latent variables can be divided into four parts: the regression coefficients ${\bm{\beta }}_i$, factor loadings ${\bm{\lambda }}_i$, the common factors ${\bf{f}}_t$, and the latent variables $h_{it}$ and $q_{jt}$ for the evolution of stochastic volatility. The four parts are dependent \emph{a posteriori}. The conditional posterior distribution is specified for each block, given the samples from the other blocks. The detailed process of prior and posterior specification can be seen in Fang \& Zhang (2014).

\subsubsection{Prior and posterior specification for ${\bm{\beta }}_i$}

The joint prior distribution of ${\bm{\beta }}_i$ $(i = 1,2, \cdots ,N)$ can be specified as a multivariate normal distribution given as
\begin{equation}\label{eq8}
{\bm{\beta }}_i  \sim N({\bm{\mu }}_\beta  ,{\bf{V}}_\beta  ),
\end{equation}
with the hyper-parameters of its mean and inverse covariance matrix having prior distributions given by:
\[
{\bm{\mu }}_\beta   \sim N({\bm{\underline \mu}}_\beta  ,{\bm{\underline \Sigma}}_\beta  )
\]
\[
{\bf{V}}_\beta ^{ - 1}  \sim W({\bm{\underline \nu}}_\beta  ,{\bf{\underline V}}_\beta ^{ - 1} ).
\]
Correspondingly, the joint posterior distribution of ${\bm{\beta }}_i$ have the multivariate normal form:
\begin{equation}\label{eq9}
{\bm{\beta }}_i |{\bf{r}}_t ,{\bf{M}},{\bm{\mu }}_\beta  ,{\bf{V}}_\beta   \sim N({\bm{\bar \beta }}_i ,{\bf{\bar V}}_i ),
\end{equation}

where ${\bf{\bar V}}_i  = ({\bf{MX}}_i^{'} {\bf{X}}_i  + {\bf{V}}_\beta ^{ - 1} )^{ - 1}$, ${\bm{\bar \beta }}_i  = {\bf{\bar V}}_i ({\bf{MX}}_i^{'} {\bf{y}}_i  + {\bf{V}}_\beta ^{ - 1} {\bm{\mu }}_\beta  )$, and ${\bf{M}}$  is the model error accuracy that can directly affect the posteriors.

\subsubsection{Prior and posterior specification for ${\bm{\lambda }}_i$}

To ensure the identification of parameters, assume that the factor loading parameters ${\bm{\lambda }}_i$ form a lower triangular matrix, i. e. $\lambda _{ij}  = 0$ ( $ i < j $; $ i = 1, \cdots ,N $; $ j = 1, \cdots ,p $ ). The prior distribution of $\lambda _{ij}$ can be set as follows:
\begin{align}
\lambda _{ij} &\sim N(0,\underline C), \quad i > j\notag\\
\lambda _{ij} &\sim  N(0,\underline C){\bf{I}}(\lambda _{ii}  > 0), \quad i = 1, \cdots ,p,\notag
\end{align}
where ${\bf{I}}(\cdot)$ is the indicator function, $\underline C$ is variance of the prior distribution.

When $i \le p$, the posterior distribution of ${\bm{\lambda }}_i$ is
\begin{equation}\label{eq10}
{\bm{\lambda }}_i  \sim N({\bm{\bar \mu }}_\lambda  ,{\bf{\bar C}}_\lambda  ){\bf{I}}({\bm{\lambda }}_{ii}  > 0).
\end{equation}

When $i > p$, the posterior distribution of ${\bm{\lambda }}_i$ is
\begin{equation}\label{eq11}
{\bm{\lambda }}_i  \sim N({\bm{\bar \mu }}_\lambda  ,{\bf{\bar C}}_\lambda  ).
\end{equation}

It should be noted that the factor loading parameters ${\bm{\lambda }}_i$ need to be specified jointly with the ${\bf{f}}_t$ as follows. The algorithm is given in the appendix.

\subsubsection{Prior and posterior specification for ${\bf{f}}_t$}

The common factors ${\bf{f}}_t$ and factor loadings ${\bm{\lambda }}_i$ need to be sampled simultaneously based on their joint posterior distributions. Compared with the factor loadings, the specification of the common factors is relatively simple. The joint posterior distribution of ${\bf{f}}_t$ in the panel data factor stochastic volatility model is related to the fixed effect parameters and the terms in the volatility equation. Its conditional posterior distribution (${\bf{f}}_t |{\bm{\beta }}_i ,{\bm{\lambda }}_i ,{\bf{r}}_t ,{\bf{x}}_t ,h_{it} ,q_{jt}$) still has a multivariate normal distribution. Given the factor loadings, ${\bf{f}}_t$ is assumed to follow the $p$ - dimensional normal distribution:
\begin{equation}\label{eq12}
{\bf{f}}_t  \sim N({\bm{\mu }}_f ,{\bf{V}}_f ).
\end{equation}
In particularly, given the factor loadings ${\bm{\lambda }}_i$, the posterior distribution of ${\bf{f}}_t$ is given by
\begin{equation}\label{eq13}
{\bf{f}}_t  \sim N({\bf{G}}^{ - 1} {\bm{\lambda }}_i^{'} {\bm{\Sigma }}_*^{ - 1} {\bf{r}}_t ,{\bf{G}}^{ - 1} ),
\end{equation}
where
$$
{\bf{G}}^{ - 1}  = {\bf{I}}_p  + {\bm{\lambda }}_i^{'} {\bm{\Sigma }}_*^{ - 1} {\bm{\lambda }}_i,
$$
and ${\bm{\Sigma }}_{*}$ is determined by the identification conditional constraints by choosing an appropriate ${\bm{\Sigma }}_{*}$ to make ${\bm{\lambda }}_i^{'} {\bm{\Sigma }}_*^{ - 1} {\bm{\lambda }}_i  = {\bf{I}}_p$.

\subsubsection{Prior and posterior specification for $(\alpha _{i0} ,\alpha _{i1} )$ and $(\varphi _{j0} ,\varphi _{j1} )$}

The conditional distribution of the logarithmic error volatility vector ${\bf{h}}_t$ can be specified as an $N$-dimensional normal distribution. In particular, we set
\begin{equation}\label{eq14}
{\bf{h}}_t |{\bf{h}}_{t - 1} ,{\bm{\alpha }}_0 ,{\bm{\alpha }}_1 ,{\bm{\Sigma }}_\nu   \sim N({\bm{\mu }}_h ,{\bf{V}}_h ),
\end{equation}
where ${\bm{\Sigma }}_\nu$ is the variance-covariance matrix of the volatility equation. Since we assume that ${\bm{\Sigma }}_\nu$ and ${\bf{V}}_h$ are both diagonal matrices, that is, there is no correlation between error terms and volatility terms. The volatility equation can be simplified as $N$ independent univariate autoregressive processes conditions (Pitt and Shephard (1999), Migon, Gamerman, Lopes and Ferreira (2005)). It could be decomposed into $N$ independent dynamic linear models. If these dynamic linear models are regarded as the evolution process, the prior, prediction and posterior conditional distribution of the volatility term ${\bf{h}}_t$ at time $t$ can be expressed respectively as:
\begin{align}
p({\bf{h}}_t |\mathscr{I}_{t - 1} ) &= \int {p({\bf{h}}_t |{\bf{h}}_{t - 1} )p({\bf{h}}_{t - 1} |\mathscr{I}_{t - 1} )d} {\bf{h}}_{t - 1}\notag\\
p({\bf{r}}_t |\mathscr{I}_{t - 1} ) &= \int {p({\bf{r}}_t |{\bf{h}}_t )p({\bf{h}}_t |I_{t - 1} )d} {\bf{h}}_t\notag\\
p({\bf{h}}_t |\mathscr{I}_t ) &\propto p({\bf{h}}_t |\mathscr{I}_{t - 1} )p({\bf{r}}_t |\mathscr{I}_{t - 1} ),\notag
\end{align}
where the information set $\mathscr{I}_{t - 1}$ is composed of the data set $\{ {\bf{r}}_t ,{\bf{x}}_t \}$ and the derived parameter set $\{ {\bm{\beta }}_i ,{\bm{\lambda }}_i ,{\bf{f}}_t ,\alpha _{i0} ,\alpha _{i1} ,\varphi _{j0} ,\varphi _{j1} ,h_{it} ,q_{jt} \}$, which can be denoted as:
 $$
 \mathscr{I}_t  = \{ {\bm{\beta }}_i ,{\bm{\lambda }}_i ,{\bf{f}}_t ,\alpha _{i0} ,\alpha _{i1} ,\varphi _{j0} ,\varphi _{j1} ,h_{it} ,q_{jt} ,{\bf{r}}_t ,{\bf{x}}_t \}.
 $$

 Since the volatility equation can be decomposed into $N$ independent components, the prior distribution of the logarithmic volatility $h_{it}$ can be specified as follow:
\begin{equation}\label{eq15}
h_{it}  \sim N(\underline \mu_h ,\underline V_h ).
\end{equation}
The blocking movement method proposed by Chib et al. (2006) is adopted to the panel data factor stochastic volatility model. Denote ${\bf{h}}^T  = ({\bf{h}}_1 , \cdots ,{\bf{h}}_T ){'}$. Both the coefficient block and the logarithmic fluctuation block are required to specify the posterior conditional distribution in the blocking movement sampling method.

Given prior distribution (\ref{eq15}), for coefficient ${\bm{\alpha }}_i$ and volatility variance $\sigma _{i\eta }^2$, the corresponding posterior distribution is also normal - inverse Gamma distribution. In a hierarchical form, it can be expressed as follows:
\begin{align}\label{eq16}
({\bm{\alpha }}_i |\sigma _{i\eta }^2 ,{\bf{r}}_t ,{\bf{x}}_t ,{\bf{h}}^T ) &\sim N({\bm{\bar \alpha }}_i ,\sigma _{i\eta }^2 {\bf{\bar V}}_h )\\
(\sigma _{i\eta }^2 |{\bf{r}}_t ,{\bf{x}}_t ,{\bf{h}}^T ) &\sim IG(\bar \nu _h /2,\bar \nu _h \bar s_h^2 /2).
\end{align}

The relationship between the prior distribution and the conjugate posterior distribution of ${\bf{h}}_t$ in equation (\ref{eq14}) can be obtained via the Bayesian law.

\subsection{MCMC Algorithm for Joint Parameter Estimation}
For convenience in algorithm design, we divide the parameters of the panel data factor stochastic volatility model into several blocks for MCMC sampling. Based on the structure of the model, it is divided into three blocks. The first block contains the regression coefficients of the explanatory variables, which we refer to as the \emph{joint parameters}. The second block contains the factor decomposition terms, including the common factors and factor loadings. The third block contains the random errors and the other terms from the corresponding stochastic volatility equation. The major advantage of Bayesian inference is its ability to incorporate prior information through MCMC. Equations (\ref{eq8}) - (\ref{eq11}) provide the foundations for specifying the joint hierarchical posterior distribution of the regression coefficients of explanatory variables. In order to sample the joint parameters ${\bm{\beta }}_i$ via Gibbs sampling and the Metropolis-Hastings (M-H) algorithm, some assumptions must be made about the error accuracy involving factorization. In algorithm design for the factor decomposition and error terms, the assumption on the error accuracy is closely related to the sampling of the joint parameters ${\bm{\beta }}_i$.

Previously, it has been assumed that the variance-covariance matrix of the error terms (including the factorization part) of the panel data stochastic volatility model is ${\bf{M}}^{ - 1} {\bf{I}}_T$. This is due to the fact that the error terms of stochastic volatility model are subject to both heteroscedasticity and sequential correlation. In order to facilitate the estimation of the joint parameters ${\bm{\beta }}_i$, combining Basu \& Chib (2003) and Chib \& Greenberg (1994) , we assume that the variance-covariance matrix is $\sigma _i^2 \vartheta _i^{ - 1} {\bf{I}}_N$ on the premise of heteroscedasticity and temporal correlation. Here, $\sigma _i^2$ has the following hierarchical form:
\begin{align}
\sigma _i^2 |\delta _\sigma   &\sim IG\left( {\frac{{\nu _\sigma  }}{2},\frac{{\delta _\sigma  }}{2}} \right)\notag\\
\delta _\sigma   &\sim G\left( {\frac{{\nu _{\sigma 0} }}{2},\frac{{\delta _{\sigma 0} }}{2}} \right),\notag
\end{align}
where $\sigma _i^2$ and $\delta _\sigma$ have inverse Gamma and Gamma prior distributions, respectively. The parameters of $\delta _\sigma$, $\nu _\sigma
$ and $\nu _\vartheta$ are scale constants. Furthermore, it is assumed that the prior distribution of $\vartheta _i$ is
$$
\vartheta _i  \sim G\left( {\frac{{\nu _\vartheta  }}{2},\frac{{\nu _\vartheta  }}{2}} \right).
$$
The aforementioned covariance matrix specification captures the heteroscedasticity of the error terms very well. The sequential correlation is also determined by the parameters   $(\alpha _{i0} ,\alpha _{i1} )$ and $(\varphi _{j0} ,\varphi _{j1} )$ of the volatility equation, with the autocorrelation parameters denoted as $\phi$. Posterior sampling can be performed based on the following algorithm:\\
(1) Sampling the joint parameters from their conditional posterior given by
\begin{equation}\label{eq18}
{\bm{\beta }}_i |{\bf{r}}_t ,\{ \sigma _i \} ,\{ \vartheta _i \} ,\phi  \sim N({\bm{\beta }}_i ,{\bf{V}}_i );
\end{equation}
(2) Sampling the scale parameter of the variance covariance matrix from
$$
\vartheta _i |{\bf{r}}_t ,\{ \sigma _i \} ,{\bm{\beta }}_i ,\phi  \sim G\left( {\frac{{\nu _\vartheta   + N}}{2},\frac{{\nu _\vartheta   + \nu }}{2}} \right);
$$
(3) Sampling the variance parameter from its conditional posterior distribution given by
$$
\sigma _i |{\bf{r}}_t ,{\bm{\beta }}_i ,\{ \vartheta _i \} ,\phi  \sim IG\left(\frac{{\nu _\vartheta   + N}}{2},\frac{{\delta _\vartheta   + \delta }}{2}\right);
$$
(4) Sampling the autocorrelation parameter according to
$$
\phi |{\bf{r}}_t ,\{ \sigma _i \} ,\{ \vartheta _i \} ,{\bm{\beta }}_i  \propto p (\phi )\prod\limits_{i = 1}^N {N({\bf{r}}_i |{\bm{\beta }}_i^{'} {\bf{x}}_i  + {\bm{\lambda }}_i^{'} {\bf{f}},\sigma _i^2 \vartheta _i^{ - 1} {\bf{I}}_N )};
$$
(5) Gibbs sampling and M-H algorithm are used to iterate until convergence. Given the posterior samples from the above algorithm, the posterior samples of $\nu$ and $\delta$ can be obtained from
\begin{align}
\nu  &= \sigma _i^{ - 2} ({\bf{r}}_i  - {\bm{\beta }}_i^{'} {\bf{x}}_i  - {\bm{\lambda }}_i^{'} {\bf{f}})^{'} ({\bf{r}}_i  - {\bm{\beta }}_i^{'} {\bf{x}}_i  - {\bm{\lambda }}_i^{'} {\bf{f}})\notag\\
\delta  &= \sum\limits_{i = 1}^N {\vartheta _i ({\bf{r}}_i  - {\bm{\beta }}_i^{'} {\bf{x}}_i  - {\bm{\lambda }}_i^{'} {\bf{f}})^{'} ({\bf{r}}_i  - {\bm{\beta }}_i^{'} {\bf{x}}_i  - {\bm{\lambda }}_i^{'} {\bf{f}})}.\notag
\end{align}
The conditional posterior density function of $\pi(\phi)$ can be approximated as the multivariate $t$ distribution.
$$
\pi(\phi |{\bf{r}}_t ,\{ \sigma _i \} ,\{ \vartheta _i \} ,{\bm{\beta }}_i ) = t(\hat \phi ,{\bf{V}}_\phi  ,\nu _\phi  ),
$$
where $\hat \phi$, ${\bf{V}}_\phi$ and $\nu _\phi$ are the location (vector) parameters, scale (matrix) parameters and degrees of freedom respectively. The degree of freedom $\nu _\phi$ can be set to any constant greater than 1. The other two parameters $\hat \phi$ and ${\bf{V}}_\phi$ will be sampled using the Forward Filtering Backward Sampling (FFBS) algorithm in section 3.5 below.

In the MCMC algorithm of joint parameters ${\bf{\beta }}_i$, the parameters $\sigma _i$, $\phi$  and $\vartheta _i$ depend on the factorization results and the settings of volatility equation. Here, $\sigma _i$ is a parameter that reflects heteroscedasticity, and $\phi$  and $\vartheta _i$ are autocorrelation parameters. Therefore, in the M-H algorithm for these parameters, it is necessary to combine the volatility equation and factor decomposition results within each iteration. This is the main difference between the panel data factor stochastic volatility model and the individual random effect panel data model in parameter estimation.

\subsection{MCMC Algorithm for Factor Decomposition}
The conditional posterior distributions of the factor loadings and common factors (factor scores) are specified as normal distributions. Based on the Bayesian law, the Gibbs sampler is used to sample the parameters with conjugate priors. The conditional posterior distribution of the factor loadings ${\bf{\lambda }}_i$ is
$$
\begin{aligned}
\pi ({\bm{\lambda }}_i |{\bf{r}}_t ,{\bf{x}}_t ,{\bm{\beta }}_i ,{\bf{f}}_t ,h_{it} ,q_{jt} ) & \propto p({\bm{\lambda }}_i )\prod\limits_{t = 1}^T {p({\bf{r}}_t |{\bm{\lambda }}_i ,{\bm{\beta }},{\bf{f}}_t ,h_{it} ,q_{jt} )} \\
& \propto p({\bm{\lambda }}_i )\prod\limits_{t = 1}^T {N({\bm{\beta }}^{'} {\bf{x}}_t ,{\bm{\Omega }}_t )},
\end{aligned}
$$
where ${\bm{\Omega }}_t  = {\bm{\Lambda Q}}_t {\bm{\Lambda }}^{'}  + {\bm{\Sigma }}_t$ is given by Equation (\ref{eq6}), and ${\bm{\Sigma }}_t  = diag(\sigma _{1t}^2 , \cdots ,\sigma _{Nt}^2 )$. The decomposition of common factors can be carried out simultaneously with the sampling of the joint parameters, with the remaining part of ${\bf{r}}_t  - {\bm{\hat \beta }}^{'} {\bf{x}}_t$ consisting of the factorization terms and the stochastic volatility terms. When the identification constraints are applied, the number of factor loadings ${\bf{\lambda }}_i
$ is $Np + p(1 - p)/2$. This causes a rather high-dimensional issue for parameter estimation, especially when the sample size is very large.

Chib, Nardari and Shephard (2006) proposed to use the t-distribution as the approximate distribution of $p({\bf{\lambda }}_i )$ as alternative distributions of the factor loadings. This method is simpler than the normal distribution approach and is easier to deal with in high-dimensional problems. The influence of the fixed effects on the parameter estimation needs to be considered in the estimation of factor loadings of the panel factor stochastic volatility. Here, we assume that the factor loadings ${\bf{\lambda }}_i$ are subject to the multivariate t-distribution.
$$
T({\bm{\lambda }}_i |s,\Xi ,\upsilon ).
$$
Denote $\ell  = \log \left\{ {\prod\nolimits_{t = 1}^T {N({\bm{\beta }}^{'} {\bf{x}}_t ,{\bm{\Omega }}_t )} } \right\}$. The location parameter $s$ in the multivariate t-distribution is generally obtained by empirical approximation on the mode of logarithms of multivariate densities. The degree of freedom $\upsilon$ can be set to any constant. The scale parameter $\Xi$ is the inverse of the second derivative of $\ell$, i.e., ${{ - \partial ^2 \ell } \mathord{\left/
 {\vphantom {{ - \partial ^2 \ell } {\partial ^2 {\bf{\lambda }}}}} \right.
 \kern-\nulldelimiterspace} {\partial ^2 {\bf{\lambda }}}}_i
$. Theoretically, approximation using the multivariate t-distribution works well with the multivariate normal distribution. Such approximation makes it easy to design relevant algorithms. The Newton-Raphson algorithm is usually used in the calculation of $s$ and $\Xi$. When dealing with higher dimensional problems, the operation time can be increased. The MCMC algorithm of factor loadings assumes subject to the multivariate normal distribution form. Since the M-H algorithm does not require symmetry in its jumping distribution in here, the multivariate t-distributions are used as the jumping distribution.

The sampling process of the common factors ${\bf{f}}_t$ is very similar to that for factor loadings. Based on a multiplicative relation among them, the factor decomposition process of the factor stochastic volatility model captures the unobservable factors of the response variables. The process combines the sampling of the joint and the influence of the volatility equation on the specification of the random error components. The idiosyncratic variance part $\sigma _\lambda ^2$ of factor decomposition is the filtering of stochastic volatility terms.

When $i \le p$, the algorithm of free elements $\lambda _{ij}$ in the factor loading matrix and the corresponding common factors (scores) is based on the following steps of sampling steps.\\
(1) Sampling $\bm{\lambda }_i$ according to the conditional posterior distribution
\begin{equation}\label{eq19}
{\bm{\lambda }}_i |{\bf{r}}_t ,{\bf{x}}_t ,{\bm{\beta }}_i ,{\bf{f}}_t ,\sigma _\lambda   \sim N({\bf{\mu }}_\lambda  ,{\bf{C}}_\lambda  ){\bf{I}}({\bm{\lambda }}_{ii}  > 0);
\end{equation}
(2) Sampling $\sigma _\lambda$ from
$$
\sigma _\lambda  |{\bm{\lambda }}_i ,{\bf{r}}_t ,{\bf{x}}_t ,{\bm{\beta }}_i ,{\bf{f}}_t  \sim IG\left( {\frac{{\nu _\lambda   + T}}{2},\frac{{\nu _\lambda  s^2  + \delta _\lambda  }}{2}} \right);
$$
(3) The M-H algorithm is used to sample the free elements of ${\bf{\lambda }}_i^*$, based either on the multivariate normal distribution or the approximate multivariate t-distribution given samples of the other parameters. The current value ${\bm{\lambda }}_i^{t - 1}$ is chose by the following rules:
$$
\pi({\bm{\lambda }}_i^{t - 1} ,{\bm{\lambda }}_i^* |{\bf{r}}_t ,{\bf{x}}_t ,{\bm{\beta }}_i ,{\bf{f}}_t ,\sigma _\lambda  ) = \min \left\{ {1,\frac{{p({\bm{\lambda }}_i^* )\prod\limits_{t = 1}^T {N({\bm{\beta }}^{'} {\bf{x}}_t ,{\bm{\Lambda }}^* {\bf{Q}}_t {\bm{\Lambda }}^{*'}  + {\bm{\Sigma }}_t^* )T({\bm{\lambda }}_i^{t - 1} |s,\Xi ,\upsilon )} }}{{p({\bm{\lambda }}_i^{t - 1} )\prod\limits_{t = 1}^T {N({\bf{\beta }}^{'} {\bf{x}}_t ,{\bm{\Lambda Q}}_t {\bm{\Lambda }}^{'}  + {\bm{\Sigma }}_t^* )T({\bm{\lambda }}_i^{*} |s,\Xi ,\upsilon )} }}} \right\},
$$
with the new free elements ${\bm{\lambda }}_i^*$ generated based on above probability. If the value is rejected, the current value ${\bm{\lambda }}_i^{t - 1}$ is accepted as a node element of the Markov chain, iterating until the stationary distribution of each $\lambda _{ij}$ is obtained.\\
(4) According to the multiplicative form of the factor decomposition, the common factors are sampled from the following distribution:
\begin{equation}\label{eq20}
{\bf{f}}_t  \sim N({\bf{G}}^{ - 1} {\bm{\lambda }}_i^{'} {\bm{\Sigma }}_*^{ - 1} {\bf{r}}_t ,{\bf{G}}^{ - 1} );
\end{equation}
(5) The algorithm iterates through the above steps until the Markov chain converges.

When $i > p$, sampling algorithms for factor loading ${\bm{\lambda }}_i$ and common factor ${\bf{f}}_t$ can be set similarly. Compared with  $i \le p$, the sampling distribution dimension of factor loading and common factor are different. For the high-dimensional factor model, the dimension of factor decomposition is decided by the purpose of dimensionality reduction.

Here, we focus on the dimension reduction regarding the number of individuals. How to reduce the dimension of individuals to an appropriate number of common factors and factor loadings depends not only on the criteria for the selection, but also on the real application of concern. Note that the selection of the number of common factors can also impact the performance of the algorithm.

\subsection{FFBS Estimation of Dynamic Stochastic Volatility Equation}
The panel data factor stochastic volatility model includes two volatility equations, namely the factor volatility equation and error volatility equation. Unlike the specification of major multivariate stochastic volatility models, the random disturbance terms of these two volatility equations are assumed to have no individual correlation. Hence, they are not expressed as the product of two independent components. In a stochastic volatility model without considering the leverage effect and jump, the volatility equation can be regarded as the state-space equation. The logarithmic $\chi ^2$ distribution can be converted into seven normal distributions of independent components. The volatility equation can be expressed as a Gaussian state-space model. The Kalman filtering algorithm of state-space model can be used for one-step forward prediction. Here, we mainly consider the Gibbs sampling algorithm proposed by Carter and Kohn (1994) and the Forward Filtering Backward Sampling (FFBS) method proposed by Fruhwirth-Schnatter (1994). The two volatility equations are both AR (1) processes of volatility terms. For the joint sampling of the two blocks, we will reform models (\ref{eq2.2})-(\ref{eq2.4}) into an alternative form involving the following two steps. Firstly, after getting the joint parameter sampling according to (\ref{eq18}) in Section 3.3, ${\bm{\hat \beta }}_i^{'} {\bf{x}}_{it}$, the observable fixed design part at the right side of model (\ref{eq2.3}) can be moved to the left side for performing logarithmic transformations. Secondly, combining models (\ref{eq2.2}) and (\ref{eq2.3}) into one model. In order to differentiate its different sources, the first $N$ terms represent error disturbance terms, and the last $p$ terms represent factor volatility terms. Let ${\bf{h}}_{t - 1}^*  = ({\bf{I}},{\bf{h}}_{t - 1} )$, ${\bf{h}}_t  = (h_{1t} , \cdots ,h_{Mt} ){'}$ be an vector with $N + p$ dimension. We can write
\begin{align}\label{eq21}
{\bf{z}}_t  &= {\bf{c}}_t  + {\bf{b}}_i {\bf{h}}_t  + {\bf{e}}_t\\
\label{eq22}
{\bf{h}}_t  &= {\bf{\alpha h}}_{t - 1}^*  + {\bm{\nu }}_t,
\end{align}
where equation (\ref{eq21}) is a transformation of equation (\ref{eq2.3}), equation (\ref{eq22}) is combined from equations (\ref{eq2.2}) and (\ref{eq2.4}), ${\bf{c}}_t$ is the drift term, and $E({\bf{e}}_t ) = {\bf{0}}$. According to the seven-component decomposition from Chib et al. (2002),  ${\bf{e}}_t  \sim N({\bf{0}},{\bm{\Sigma }}_e )$. Amongst the volatility equation (\ref{eq22}), the dynamic coefficients are ${\bm{\alpha }} = (\alpha _{i0} ,\alpha _{i1} )'{\bf{I}}_{i \le N}  + (\varphi _{i0} ,\varphi _{i1} )'{\bf{I}}_{N < i \le N + p}$, ${\bm{\nu }}_t  \sim N({\bf{0}},{\bm{\Sigma }}_\nu  )$.

Models (\ref{eq21}) and (\ref{eq22}) convert the factor panel stochastic volatility model into the special Gaussian dynamic linear model. In addition to the observable part ${\bf{z}}_t$, we sample the volatility coefficients ${\bm{\alpha }}$ through the particle filtering algorithm so as to extrapolate the prediction of the latent variables ${\bf{h}}_t$. The FFBS block sampling algorithm needs to construct Markov chains to extract discrete samples from a block which has been set beforehand. Here, $\mathscr{I}_t  = \mathscr{I}_{t - 1}  \cup \{ {\bf{z}}_t \}$ is the information set up to time $t-1$ plus the observation values of the explanatory variables and the response variable at time $ t $. These values are generally observable. ${\bf{I}}_t$ contains the full observable information set and the unobservable information set at time $t$. Given the other parameters and sample data, ${\bf{h}}$ is denoted as an implied volatility block, of which the joint full conditional posterior distribution is $\pi ({\bf{h}}|{\bm{\Sigma }}_e ,{\bm{\Sigma }}_\nu  ,{\bf{I}}_T )$. The conditional distribution is characterized by the volatility equation and the generation process of the latent variables:
\begin{equation}\label{eq23}
\pi ({\bf{h}}|{\bm{\Sigma }}_e ,{\bm{\Sigma }}_\nu  ,{\bf{I}}_T ) = p({\bf{h}}_T |{\bm{\Sigma }}_e ,{\bm{\Sigma }}_\nu  ,{\bf{I}}_T )\prod\limits_{t = 1}^T {p({\bf{h}}_t |{\bf{h}}_{t + 1} ,{\bm{\Sigma }}_e ,{\bm{\Sigma }}_\nu  ,\mathscr{I}_t )}.
\end{equation}
The joint posterior distribution of the implied volatility blocks are determined by the conditional distribution at time $T$. Therefore, the conditional distribution can be described as:
\begin{equation}\label{eq24}
\begin{aligned}
p({\bf{h}}|{\bm{\Sigma }}_e ,{\bm{\Sigma }}_\nu  ,\mathscr{I}_T ) & = p({\bf{h}}_2 , \cdots ,{\bf{h}}_T |{\bm{\Sigma }}_e ,{\bm{\Sigma }}_\nu  ,\mathscr{I}_T )\\
 & = p({\bf{h}}_T |{\bm{\Sigma }}_e ,{\bm{\Sigma }}_\nu  ,\mathscr{I}_T )p({\bf{h}}_{T - 1} |{\bf{h}}_T ,{\bm{\Sigma }}_e ,{\bm{\Sigma }}_\nu  ,I_T ) \cdots p({\bf{h}}_1 |{\bf{h}}_2 , \cdots ,{\bf{h}}_T ,{\bm{\Sigma }}_e ,{\bm{\Sigma }}_\nu  ,\mathscr{I}_T)\\
 & = p({\bf{h}}_T |{\bm{\Sigma }}_e ,{\bm{\Sigma }}_\nu  ,\mathscr{I}_T )p({\bf{h}}_{T - 1} |{\bf{h}}_T ,{\bm{\Sigma }}_e ,{\bm{\Sigma }}_\nu  ,I_T ) \cdots p({\bf{h}}_1 |{\bf{h}}_2 ,{\bm{\Sigma }}_e ,{\bm{\Sigma }}_\nu  ,\mathscr{I}_T ).
\end{aligned}
\end{equation}
The last step of equation (\ref{eq24}) is obtained from the backward property of the Markov chain. Here, ${\bf{h}}_t$ is conditionally independent of ${\bf{h}}_{t + j}(j\geq1)$.

According to Bayesian law, the conditional distribution of $({\bf{h}}_t |{\bf{h}}_{t + 1} ,{\bm{\Sigma }}_e ,{\bm{\Sigma }}_\nu  ,\mathscr{I}_t )$ in formula (\ref{eq24}) can be acquired from the conditional transition probability function $p({\bf{h}}_{t + 1} |{\bf{h}}_t ,{\bm{\Sigma }}_e ,{\bm{\Sigma }}_\nu  ,\mathscr{I}_t )$ and the conditional probability function $p({\bf{h}}_t |{\bm{\Sigma }}_e ,{\bm{\Sigma }}_\nu  ,\mathscr{I}_t )$. Denote $E({\bf{h}}_t |\mathscr{I}_T ) = {\bf{m}}_t$ and $Var({\bf{h}}_t |\mathscr{I}_T ) = {\bf{D}}_t$. We have
\begin{equation}\label{eq25}
({\bf{h}}_t |{\bf{h}}_{t + 1} ,{\bm{\Sigma }}_e ,{\bm{\Sigma }}_\nu  ,\mathscr{I}_t ) \sim N({\bm{\mu }}_h^* ,{\bf{V}}_h^* ),
\end{equation}
where
\begin{align}
{\bm{\mu }}_h^*  &= ({\bm{\alpha }}^{'} {\bm{\Sigma }}_\nu ^{ - 1} {\bm{\alpha }} + {\bf{m}}_t^{ - 1} )^{ - 1} ({\bm{\alpha }}^{'} {\bm{\Sigma }}_\nu ^{ - 1} {\bf{h}}_{t + 1}  + {\bf{D}}_t )\notag\\
{\bf{V}}_h^*  &= ({\bm{\alpha }}^{'} {\bm{\Sigma }}_\nu ^{ - 1} {\bm{\alpha }} + {\bf{m}}_t^{ - 1} )^{ - 1}\notag.
\end{align}
The joint sampling of ${\bf{h}}_t$ consists of two steps: backward sampling and forward Kalman filtering. Note that $({\bf{h}}_t |\mathscr{I}_T )$ is subject to the normal distributions. Using the Kalman filtering algorithm, the estimation of the mean ${\bf{m}}_t$ and variance   ${\bf{D}}_t$ of $({\bf{h}}_t |\mathscr{I}_T )$ can be obtained. Since ${\bf{h}}_{t - 1}$ and ${\bf{I}}_t$ are independent to each other, from Markov chain $({\bf{h}}_{t - 1} |{\bf{h}}_t )$,  we have
\[
p({\bf{h}}_{t - 1} |{\bf{h}}_t ,{\bm{\Sigma }}_e ,{\bm{\Sigma }}_\nu  ,I_t ) = p({\bf{h}}_{t - 1} |{\bf{h}}_t ,{\bm{\Sigma }}_e ,{\bm{\Sigma }}_\nu  ,I_{t - 1} ),
\]
where $p({\bf{h}}_{t - 1} |{\bf{h}}_t ,{\bm{\Sigma }}_e ,{\bm{\Sigma }}_\nu  ,\mathscr{I}_t )$ can be sampled from the information set $\mathscr{I}_{t - 1}$ via the mean equation and the volatility equation. The implied volatility block ${\bf{h}}$ is sampled in MCMC. The sampling algorithm of the implied volatility blocks of the panel data factor stochastic volatility model can constructed via the multivariate FFBS as follows:

(1) Using Kalman filtering to filter ${\bf{h}}_t$ from the conditional probability density $p({\bf{h}}_t |{\bm{\Sigma }}_e ,{\bm{\Sigma }}_\nu  ,I_t )$, forming a series of parameters and the probability density function of ${\bf{h}}_t$ ( $t = 1, \cdots ,T - 1$).

(2) The current value ${\bf{h}}_t$ of the state vector is sampled from the distribution given by the marginal density (\ref{eq25}).

(3) The previous value ${\bf{h}}_{t - 1}$ is sampled from the conditional probability $p({\bf{h}}_{t - 1} |{\bf{h}}_t ,{\bm{\Sigma }}_e ,{\bm{\Sigma }}_\nu  ,\mathscr{I}_{t - 1} )$ by backward sampling. Return to step (2) until $t = 1$ and complete the backward sampling process.

In all, the FFBS method includes two steps: forward filtering and backward sampling. The algorithm we propose can take the advantage of the information until time $t$ to predict ${\bf{h}}_t$. The unobservable volatility terms ${\bf{h}}_t$ have sequential correlation, a situation for which the two-step algorithm is more effective. The multivariate sampling algorithm is also favorable for capturing the correlation between individuals and for simultaneously accounting for the pairwise correlations of the panel data sequence and the cross sectional characteristics. In the multivariate FFBS, both the conditional probability density and marginal probability density of latent variables ${\bf{h}}_t$ satisfy the multivariate normal distribution. In the joint sampling for the volatility equation with multiple individuals, the correlation between individuals is captured by the parameters of multivariate normal distribution given in equation (\ref{eq22}).

\section{\sc {Simulation Studies}}
The posterior estimation of PDFSVM is a joint estimation process. The three-block sampling method proposed in here captures the aforementioned structural characteristics of each component. As mentioned earlier, the full model is divided into three blocks in the prior and posterior specification: the joint parameters, factor decomposition and volatility equation whilst determining the posterior distributions for constructing the MCMC algorithm. In the actual sampling process, these three blocks of parameters are sampled sequentially and cannot be separated completely. From the joint parameter estimation, the influence of the observable factors on the response variables can be assessed according to the estimation of regression coefficients. The factorization results of unobservable factors can be used to explain the unobservable stochastic volatility factors. Thus, the factorization process should be combined with the estimation of the stochastic equation. In order to validate the effectiveness of the joint estimation method designed in the previous section, we consider the influence of unobservable factors firstly, then add the observable variables to estimate the mean equation.

From the perspective of model fitting and efficiency, we mainly discuss four aspects: the estimation accuracy, the robustness of the estimation results, the prior distribution, and the influence of the initial values on the estimation results. In the simulation study, a set of models are constructed to mimic real application problems. For example, we consider a large portfolio consisting of 20 to 40 financial assets. It includes not only observable factors but also unobservable factors on the price changes of financial assets. In order to study the estimation efficiency of joint estimation based on MCMC, the data generation process is designed according to models (\ref{eq2.2}) - (\ref{eq2.4}). The high-dimensional property of the panel data factor stochastic volatility model are characterized by the number of individuals $N$, the period length $T$, the number of covariates $k$, and the number of common factors $p$. In the simulation study, we consider eight different models to reflect the impact of these numbers on the estimation results. The model dimension settings are shown in table 1.

\begin{table}
\caption{Model dimensional settings for simulated data sets.}
\begin{center}
\begin{tabular} {c|ccccc||c|ccccc}\hline\hline
 Model & $N$& $k$& $p$& $T$& No. of & Model& $N$& $k$& $p$& $T$& No. of \\
                     &&&&& Parameters              && &&&& Parameters\\

\hline
M1&	10&	3&	3&	200&	80&	    M5&	10&	4&	4&	200&   98\\
M2&	20&	3&	3&	200&	160&	M6&	20&	4&	4&	200&   198\\
M3&	10&	3&	3&	400&	80&	    M7&	40&	4&	4&	400&   398\\
M4&	20&	3&	3&	400&	160&	M8&	40&	4&	6&	1000&  471\\

\hline\hline
\end{tabular}
\end{center}
\label{tab1}
\end{table}

The generation process of the response variable $r_{it}$ is determined by the high-dimensional parameters involved in each of model elements. After specifying the numbers of individuals, periods, covariates and common factors, we randomly generate the parameters and data sets including each free element $\lambda _{ij}$ of the factor loading matrix ${\bm{\lambda }}$, covariates ${\bf{x}}_{it}  = (1, \cdots ,x_{kit} )$, regression coefficients ${\bm{\beta }}_i^{'}  = (\beta _{i1} , \cdots ,\beta _{ik} )$, factor volatility coefficients $(\alpha _{i0} ,\alpha _{i1} )$, stochastic volatility coefficients $(\varphi _{j0} ,\varphi _{j1} )$, random error terms $v_{it}$ and $w_{jt}$. It is assumed that the parameters are independent of each other except for the correlations inherited from the model setting.

The data generation process (DGP) for the scenarios in table 1 are set as follows:

(1) Generate ${\bf{x}}_{it}$ (a total of $k - 1$ variables in addition to the constant terms), assuming that the observable factors $x_{2it}$, $ \cdots$, and $x_{kit}$ all come from different normal distributions. For example, the $a$-th explanatory variable $x_{ait}$ is generated by a normal distribution with mean $2a$ and variance $2^a$ ($a = 2, \cdots ,k$ ).

(2) Generate ${\bm{\beta }}_i^{'}$ for the random intercept model where the coefficient of each explanatory variable changes with the individual. For individual $i$, for example, the coefficients $\beta _{i1}$, $\cdots$ , and $\beta _{ik}$ are obtained from the normal distribution. Without loss of generality, its mean and variance are assumed to be 0.06 and 0.009.

(3) Generate $\lambda _{ij}$ from normal distributions. For any $i = 1,2, \cdots, N$; $j = 1,2, \cdots ,p$, we assumed $\lambda _{ij}  \sim N(0.8,0.1)$.

(4) Generate $(\alpha _{i0} ,\alpha _{i1} )$. For any $i = 1,2, \cdots ,N$, we assume $\alpha _{i0}  \sim N(0.08,0.01)$, with $\alpha _{i1}$ generated from a rescaled Beta distribution with mean 0.85 and variance 0.25.

(5) Generate $(\varphi _{j0} ,\varphi _{j1} )$. For any $j = 1,2, \cdots ,p$, we assume $\varphi _{j0}  \sim N(0.09,0.01)$, with $\varphi _{j1}$ generated from a rescaled Beta distribution with mean 0.95 and variance 0.3.

(6) Generate $v_{it}$ from the standard normal distribution $N(0,1)$, assuming the error term is white noise.

(7) Generate $w_{jt}$ from the standard normal distribution $N(0,1)$, assuming the error term is white noise.

(8) Set the initial value of $h_{it - 1}$ to 0.0.

(9) Set the initial value of $q_{jt - 1}$ to 0.0.

Note that the generation process of the random disturbance terms and factor volatility terms depends not only on $h_{it}$ and $q_{jt}$ that are generated by volatility equation (\ref{eq2.2}) and (\ref{eq2.3}), but also on its hierarchical form.

(10) Generate $u_{it}$ according to the setting of random error terms, with $u_{it}  = \exp (h_{it} /2)\eta _{it}$, where $\eta _{it}  \sim N(0,1)$.

(11) Generate the factor volatility terms ${\bf{f}}_t$ with $f_{jt}  = \exp (q_{jt} /2)\varepsilon _{jt}$, where $\varepsilon _{jt}  \sim N(0,1)$.

From steps (1) - (11) of the DGP, multivariate time series $\{ r_{it} \} (i = 1, \cdots ,N)$ with length $T$ can be generated. Our purpose is to establish the panel data factor stochastic volatility model for applying the aforementioned MCMC algorithm for model esitmation, and validate the estimation efficiency and performance. For the prior specification of the high-dimensional factor model, we assume independent prior distributions, in order to improve the algorithm performance. The prior specifications are as follows:

The prior distribution of the joint parameters is $\beta _{ij}  \sim N(0.02,0.04)$. We neglect the difference between sampling by column and sampling by row here. The free elements of factor loadings $\lambda _{ij}  \sim N(0.9,0.1)$. $\alpha _{i0}$  and $\varphi _{j0}$ are assumed to have the normal distribution $N( - 0.04,0.01)$. From Chib et al. (2006), it is assumed that $\alpha _{i1}  = 2\alpha _{i1}^*  - 1$ and $\varphi _{j1}  = 2\varphi _{j1}^*  - 1$, where $\alpha _{i1}^*  \sim Beta(0.85,0.2)$  and $\varphi _{j1}^*  \sim Beta(0.9,0.25)$. For cases involving degrees of freedom, we assume that the degrees of freedom come from the uniform distribution of lattice points (7,10,13,16,19,22,25,30). Prior and posterior specification of the other parameters are given in equations (\ref{eq2.2}) - (\ref{eq2.3}).

In posterior sampling, we consider how to combine the factor decomposition and FFBS algorithm, after considering the interaction of each system of equations and the others. Whereas the factorization process is an in-depth analysis of unobservable factors, the FFBS algorithm takes into account the generation process of the latent variables. Hence, the filtering process in the forward prediction is closely related to the factor decomposition process. Furthermore, there is a nesting structure between the potential fluctuation terms $h_{it}$, the factor fluctuation terms $q_{jt}$ and the estimation of the joint parameters. Based on the relationships between the three, the follow steps must be obeyed: (1) sampling observable factors before the non-observable factors and (2) carrying out component decomposition before dynamic linear model estimation. Due to various factors that need to be considered, the overall process of model implementation is very complex. Here, we mainly focus on the estimation of the regression coefficients and other parameters that can describe the fluctuation characteristics of explanatory variables. Without loss of generity, we generate 12,000 posterior samples for each parameter, with the first 2,000 samples are discarded as burn-in and the last 10,000 samples are retained for inference.

The three blocks of parameters from the panel data factor stochastic volatility model need to be sampled jointly in each iteration, as the posterior distribution of each block depends on the current values of the other two blocks. In order to verify the estimation performance of the proposed algorithm, we consider the four parts of parameters and random terms: the factor loadings, the implied volatility terms, the scores of the common factors, and the regression coefficients of the covariates. The implied volatility terms include the factor volatility and random volatility terms. In what as follows, we only report the simulation results from model $M1$. The results of other models are similar, and will be reported only when a comparison needs to be done between the models.
\begin{figure}[ht]
\includegraphics[bb=35 500 15 730,scale=.9]{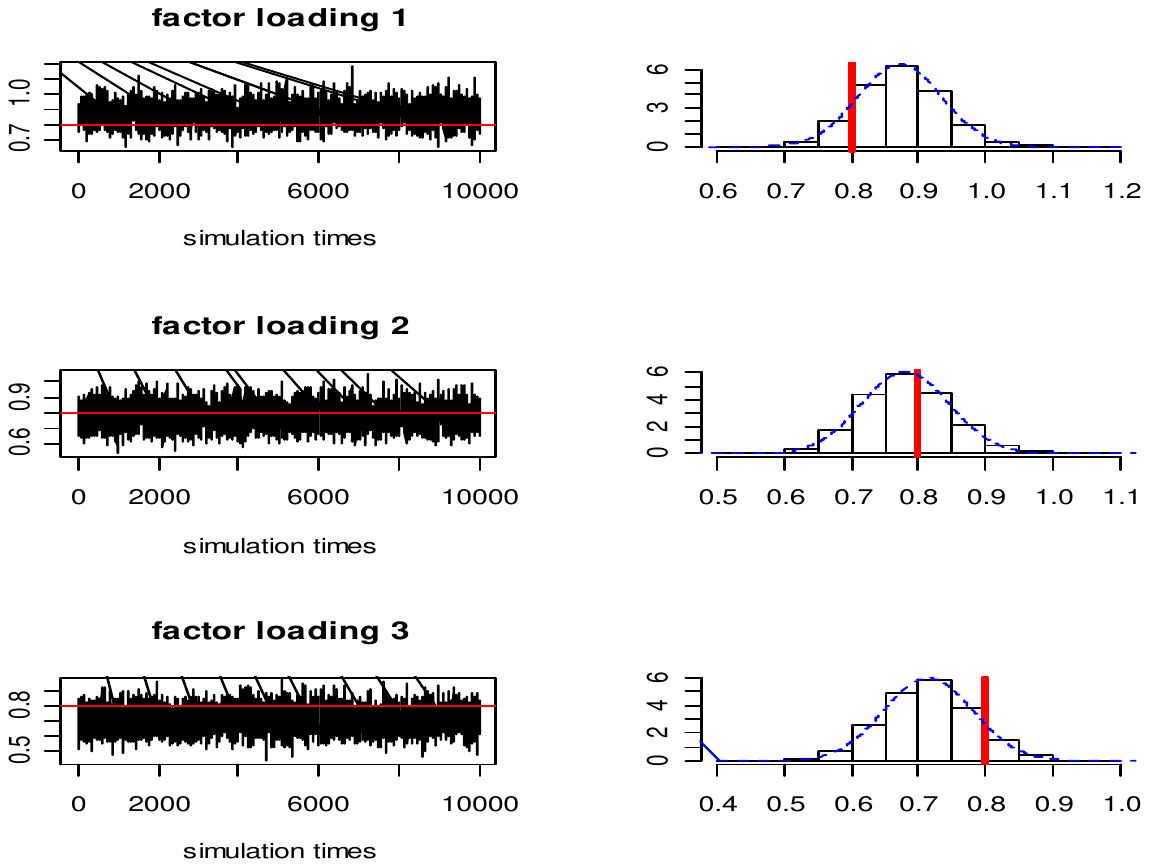}
\caption{Posterior samples of factor loadings for the three factors of Model $M1$.}
\end{figure}

The sampling of the factor loadings $\lambda _{ij}$ depends on the choice of the number of factors. In simulation model $M1$, the number of individuals is 10 and the observation period is 200. According to the ICp criterion proposed by Bai and Ng (2002), the number of common factors is 3, so the factor loading matrix ${\bm{\Lambda }}$ contains 10 rows and 3 columns. Based on the 10,000 posterior samples, the average factor loading corresponding to the three common factors are calculated respectively. From figure 2, after 2000 simulations, the distribution of the posterior samples of the factor loading tends to be stationary, with the posterior distribution covering the mean of the initial distribution. The histograms show that the posterior distributions are approximately normal. Compared with the initial value used in the data generation process, the location parameter has some deviation. Such deviation can be explored in the future research. The posterior samples of the factor loadings can be generated simultaneously from the above MCMC algorithm.

With the generation mechanism, the sampling of the stochastic volatility terms ${\bf{h}}_t$ and factor volatility terms ${\bf{q}}_t$ are related to the factor decomposition process and the state transition rule of the volatility equation at the same time. In the FFBS algorithm, the filtering and sampling the two groups of latent state variables are apparently different due to the structural difference in the models. The estimates of the stochastic volatility terms ${\bf{h}}_t$ describe the heteroscedasticity and stochastic volatility of the model, while the possible correlation in the sequence of error terms reflects the error structures of the mean equation. The estimates of the factor volatility terms ${\bf{q}}_t$ describe the structure characteristics of the common factors. The common factors obtained by factorization are latent variables, with the estimate reflecting the impact of each common factor. For the simulation study, the estimates and the true values of the factor scores can be compared for assessing the effectiveness of factor volatility model and factor decomposition process.

\begin{figure}[htbp]
\includegraphics[bb=60 450 10 700,scale=1]{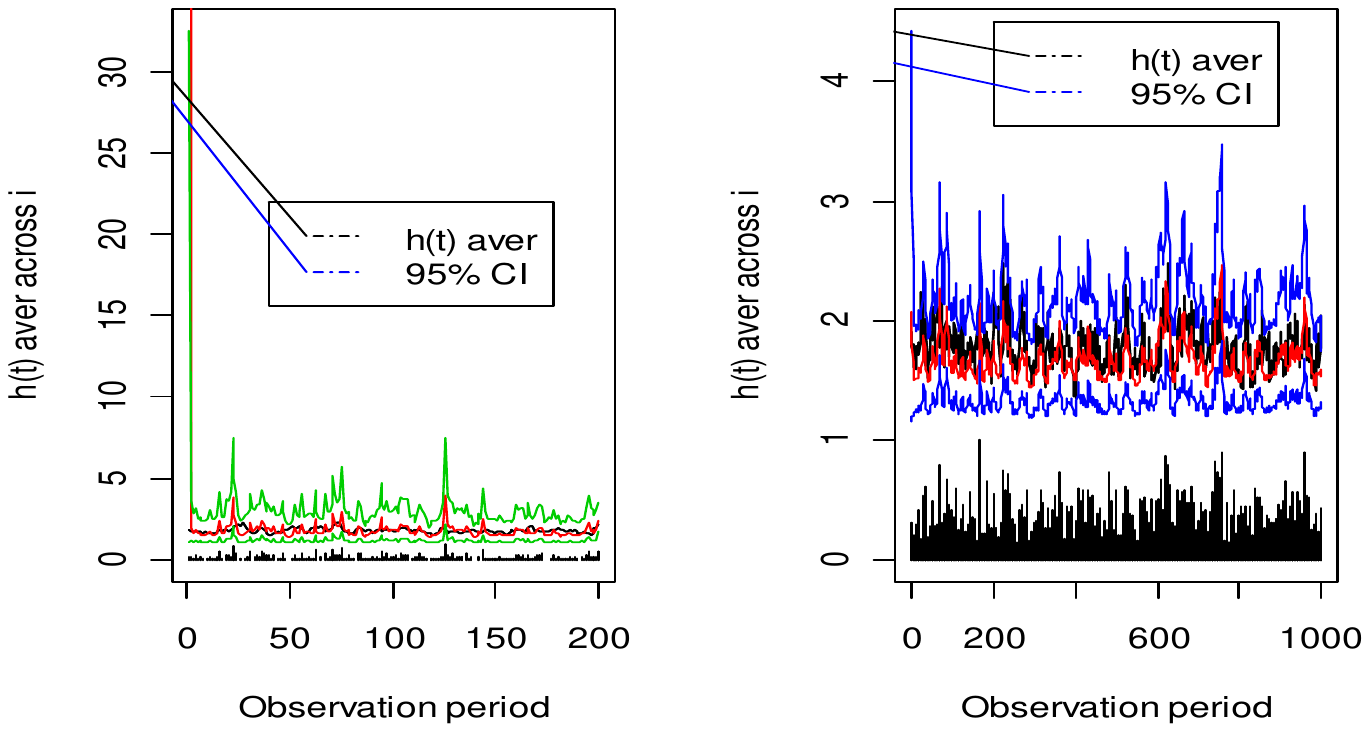}
\caption{The latent factors estimated by multivariate FFBS joint estimation. }
\end{figure}

The sampling process is similar to the two types of latent factors. Figure 3 shows the posterior samples of the latent factors of random error terms. In order to reflect the change of latent factors over time, we average the latent factors for each individual. The observation period in the left figure is $T=200$, and the right figure is 1000. Obviously, with the growth of observation period, the posterior samples of the latent factors tend to stationary distributions. This shows that the FFBS method for the panel data factor stochastic volatility model works better for data with a long observation period. This is consistent with the requirement of a long observation period by the mixed double factor model and the discrete panel data dynamic factor model, according with the large number of outliers at the beginning of initial observation periods. When the forward filtering and backward sampling algorithm is adopted, the outliers in the initial stage are effectively accounted for later in the observation period. This is the reason that the performance of the model seems well in the left and right panels of the figure. The posterior mean of the latent factors (represented by ``$h(t)\,aver$" in the figure) almost coincides with the true value, both falling within the 95$\%$ confidence interval, which shows the joint sampling based on FFBS captures the transfer rule of the potential factors better. In figure 3, the scatter plots below the two graphs are based on the posterior samples of the variance of the latent factors. For the convenience of comparison, we used a different scale (raised by one unit) for the posterior samples of the mean of the latent factors. It can be seen that the errors of the latent factors have apparent clustering characteristics, revealing the heteroscedasticity of the stochastic volatility model.

For the latent factor terms, the scores of common factors and the random error components of the mean equation can be obtained through the relationship between the generation process of common factors and latent factors expressed as follows:
$$
f_{jt}  = \exp (q_{jt} /2)\varepsilon _{jt}.
$$
In order to compare the true factor scores and their estimates, the individual average method is used to obtain the estimates of the common factor scores. Let $\bar f_t$ represents the common factor individual average score. We can write
$$
\bar f_t  = \sum\limits_{j = 1}^p {\exp (q_{jt} /2)\varepsilon _{jt} }.
$$
The individual average of the true factor scores is calculated in the same way as that for the posterior estimates, with the comparison, the two results are given in figure 4. The left panel of figure 4 shows the fitting performance based on the true factor scores and estimated scores based on 200 observations. The middle panel gives the scatter plot of the true factor scores (x-axis) versus the fitted (y-axis). In the middle panel, the observations closely scattered around the diagonal line, which indicates favorable fitting performance. The right panel shows the autocorrelogram of the posterior samples of the factor scores. With the AR (1) process added to the stochastic volatility effect, there is no obvious high-order autocorrelation of the factor score items. Overall, the results of figure 4 shows that the joint method of filtering and sampling is effective in estimation the latent factor volatility.

\begin{figure}[ht]
\includegraphics[bb=95 550 15 700,scale=1.2]{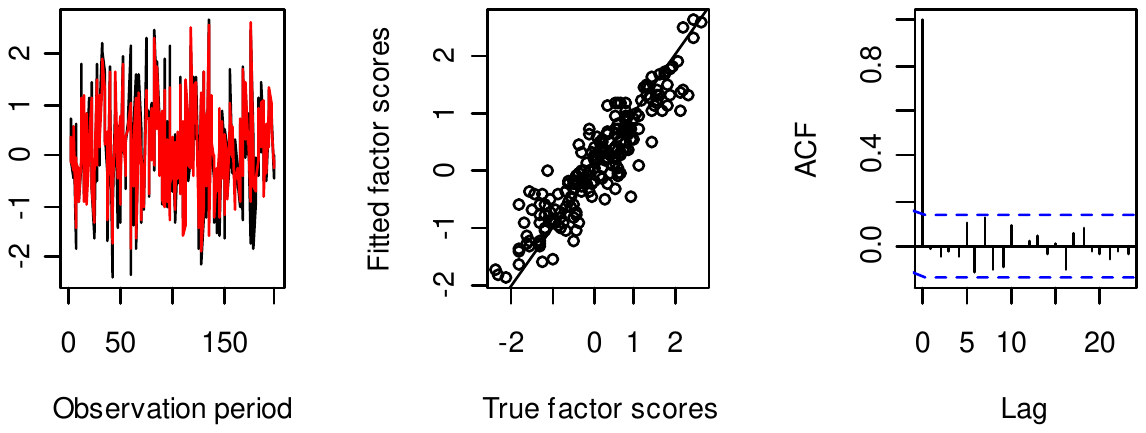}
\caption{Comparison of the true and fitted values of the common factor scores.}
\end{figure}

The favorable performance of the joint algorithm is reflected not only through the fitting of the latent fluctuation terms and factor error terms but also through the fitting of the interception term and coefficients of the autoregressive terms of the AR (1) process in the volatility equation. For the data generation processes, the constant term $\alpha _{i0}$ and slope term $\alpha _{i1}$ of factor volatility equation are set to 0.08 and 0.85, respectively. From the simulation study, we observe that $\alpha _{i1}$ gets closer to 1 as soon as $\alpha _{i0}$ gets closer to 0, showing the goodness of fit resulting from the assumption $E(f_{jt} ) = 0$. Combined with other conditions, the assumption $E(f_{jt} ) = 0$ help guarantee the unique identification of the factorization results. The left panel of figure 5 presents the individual average of the interception $\alpha _0$ and the slope term $\alpha _1$. The right penal presents the frequency based on the posterior samples, showing that the posterior samples oscillate around the true values and the posterior mean gradually tends to the true values of 0.08 and 0.85 after a burn-in of 2000 samples. The frequency figure validates that the mode is very close to the true value. With the increase of the chain length and observation period, the posterior samples of the coefficients converges to the true values very well.

\begin{figure}[ht]
\includegraphics[bb=95 450 15 700,scale=1]{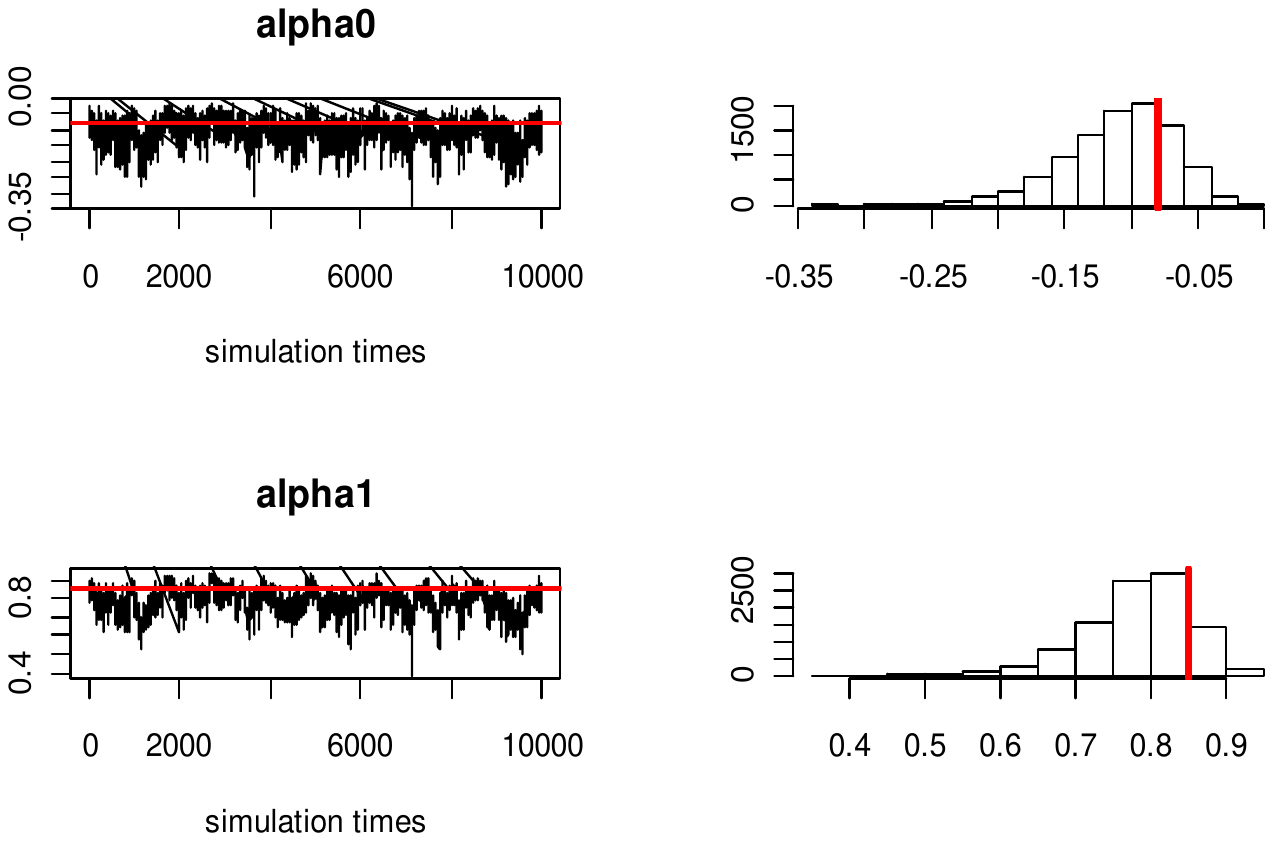}
\caption{Posterior samples of coefficients from volatility equation.}
\end{figure}

Furthermore, we verify the estimation accuracy of the regression coefficients $\beta _{ij}$. Compared with the simple stochastic volatility model and the multivariate stochastic volatility model, the regression coefficients of the factor panel data model has specific real implications. Based on such implications, we can further analyze estimation results of $\beta _{ij}$. Here, we verify the performance of joint algorithm with the FFBS algorithm for the latent fluctuation terms and hierarchical Bayesian factor decomposition. Based on the data generation process designed earlier, we use the joint algorithm to implement Models $M1$-$M8$ in table 1. Here, we only report the estimation results of models $M1$ and model $M5$. The number of individuals in model $M1$ and model $M5$ is 10, and the number of covariates is 3 and 4 respectively. Both models include interception terms. The regression coefficients $\beta _{ij}$ are generated by a normal distribution with mean 0.6 and variance 0.009.

In the panel data factor stochastic volatility model, the influence of various observable factors on the response variables is captured by the covariates and individual effects. In the joint algorithm, the regression coefficients are estimated by the hierarchical Bayesian method of the panel data random coefficient model, with each model assumed to have interception terms. In the MCMC algorithm, a total of 10,000 posterior samples are taken after the burn-in period. The simulation was repeated 1000 times to obtain the estimates of coefficients of each individual variable, according to the new data generated process. The estimated standard error of each coefficient and the root-mean-square error (RMSE) are very small, so we only report the average of the estimated coefficient and the $t$-statistic. The estimates and $t$-statistics are presented in table 2.

\begin{table}[ht]
\caption{Estimates and $t$-statistics for regression coefficients of model $M1$ and model $M5$.}
\begin{center}
\begin{tabular} {c|cccccccccc}\hline\hline


 Model1&	ind1&	ind2&	ind3&	ind4&	ind5&	ind6&	ind7&	ind8&	ind9&	ind10\\

\hline
const&	    0.058& 	0.058& 	0.057& 	0.057& 	0.057& 	0.058& 	0.057& 	0.057& 	0.057& 	0.058\\
t-value&	21.22& 	21.31& 	21.24& 	21.39& 	21.13& 	21.30& 	21.40& 	21.33& 	20.98& 	21.34\\
var1&	    0.061& 	0.060& 	0.060& 	0.060& 	0.060& 	0.060& 	0.060& 	0.060& 	0.060& 	0.061\\
t-value&   68.51& 	66.81& 	66.43& 	65.67& 	65.26& 	68.29& 	68.98& 	66.94& 	69.89& 	69.26\\
var2&	    0.056& 	0.056& 	0.055& 	0.056& 	0.055& 	0.056& 	0.055& 	0.055& 	0.056& 	0.056\\
t-value&   64.64& 	63.00& 	62.55& 	63.00& 	64.88& 	62.56& 	61.19& 	64.45& 	64.18& 	62.88\\
var3&	    0.057& 	0.058& 	0.056& 	0.057& 	0.057& 	0.058& 	0.057& 	0.057& 	0.057& 	0.058\\
t-value&   66.20& 	67.89& 	64.66& 	65.21& 	65.77& 	66.36& 	66.26& 	65.48& 	66.26& 	66.80\\
\hline
 Model5&	ind1&	ind2&	ind3&	ind4&	ind5&	ind6&	ind7&	ind8&	ind9&	ind10\\
\hline
const&	  0.059& 	0.059& 	0.059& 	0.059& 	0.059& 	0.059& 	0.059& 	0.059& 	0.059& 	0.059\\
t-value&  24.14& 	24.09& 	24.34& 	24.16& 	24.08& 	24.15& 	24.15& 	24.30& 	24.17& 	24.28\\
var1&	  0.055& 	0.054& 	0.053& 	0.054& 	0.054& 	0.054& 	0.054& 	0.054& 	0.054& 	0.054\\
t-value&  61.77& 	62.37& 	59.33& 	61.13& 	61.06& 	61.25& 	62.88& 	62.12& 	62.00& 	61.24\\
var2&	  0.058& 	0.058& 	0.058& 	0.058& 	0.058& 	0.059& 	0.058& 	0.058& 	0.058& 	0.058\\
t-value&  66.89& 	64.90& 	64.94& 	66.40& 	67.29& 	66.22& 	66.81& 	65.17& 	66.97& 	66.27\\
var3&	  0.058& 	0.058& 	0.057& 	0.057& 	0.057& 	0.058& 	0.058& 	0.058& 	0.057& 	0.058\\
t-value&  65.51& 	66.85& 	65.26& 	64.48& 	66.11& 	64.77& 	64.45& 	65.20& 	64.83& 	65.88\\
var4&	  0.058& 	0.058& 	0.057& 	0.057& 	0.057& 	0.058& 	0.057& 	0.058& 	0.057& 	0.058\\
t-value&  64.26& 	65.06& 	62.15& 	64.54& 	64.31& 	65.51& 	64.10& 	63.41& 	65.68& 	65.41\\

\hline\hline
\end{tabular}
\label{tab2}
\end{center}
\end{table}

In table 2, $ind$ represents individuals and $var$ represents variables. And $const$ represents the interception term of the model. The estimation of each coefficient is the average of the 1000 repeated simulations. The $t$-statistic value under the estimated coefficient is computed based on the average of the estimated standard error. The estimated values of coefficients from the two models are very close to the mean of the normal distribution, with the all $t$-tests all are highly significant. This shows the super performance goodness of fit. The results of numerical simulations show that the deviation between the estimated value and the true value depends on the estimation method and the initial data generation process. The estimates' result will be more accurate when we reduce the variance of the normal distribution of $\beta _{ij}$ in the data generation process.

The simulation results demonstrate that the proposed algorithm is effective for the panel data factor stochastic volatility model. The estimated coefficient of the mean equation or stochastic equation and the extraction of latent factors and random error components both have good fitting performance on simulated data. The estimates of the observable factors are reflected in the individual random coefficients, while the unobservable factors are determined by the volatility equation and factor decomposition. From the simulation studies, we can study that each part can reflect the distribution features observed in real data. The estimates from the proposed algorithm are accurate in overall, with the mean structure of $r_{it}$ being estimated more precisely. Hence, the panel data factor stochastic volatility model implemented with FFBS has the capacity of capturing the influence of observable factors and unobservable factors very well.

\section{\sc {Comparative Analysis on Internet and Traditional Financial Listed companies in China}}

Compared with other stochastic volatility models, the main advantage of the factor panel stochastic volatility model is its ability to simultaneously capture the influence of observable and unobservable factors on the investment return of financial assets. The observable and unobservable factors can be represented by influencing factors from the real world. Explanatory variables in the panel data stochastic volatility model represent influencing factors that are observable, whereas the unobservable factors are represented by latent factors and random error terms. In research concerning the stock market, observable factors can include market factors, over-the-counter market factors, and industrial factors related to the company. Unobservable factors can have the same sources as observable factors but are not observable or captured in the available data. By analyzing various factors, companies may be able to minimize the risk of financial assets, optimize the allocation of financial assets, as well improve the investment return.

In recent years, internet finance started booming in China's financial market, the concept of internet finance attract great attention in the capital market. Many listed companies have incorporated internet finance into their main businesses. Internet financial companies take advantage of modern information technology and take traditional financial businesses from the counter to the network. Here, we use the term 'internet financial company' for a listed company with main businesses including internet financial technologies, internet financial services, direct or indirect control a P2P online lending platforms fully. Traditional financial companies are those that rely on banking as the main businesses solely. P2P lending platform and crowd funding are regarded as newcomer Internet-based financial businesses in China. Many listed companies have paid great interesting in investing in internet finance. Therefore, we will compare influence of the observable and unobservable factors on the stock returns of internet financial companies and traditional financial listed companies in the Chinese stock market.

The internet financial companies have been growing rapidly in recent years in China. For comparative analysis purposes, we select 10 listed companies from the three classes of internet financial companies defined above. Because the state-owned commercial banks in China are very large, we select 10 regional commercial banks from the public banking sector. In China, regional commercial banks refer to banking financial institutions whose business area is subject to regional restrictions. All of these listed companies come from the stock markets of Shenzhen A-shares, Shanghai A-shares and the Growth Enterprise Market (GEM). For simplicity, the stock code is used to represent the name of the listed company. The trade data of the listed companies come from the CSMAR China Stock Market Trading Database. We apply the panel data factor stochastic volatility model based on the company trading records. For the model, we assume that the mean equation and common factors have volatility evolution, with both the excess logarithmic rate of return sequence and the factor component assumed to have lag effects. Due to such assumptions, the continuity of data period must be ensured. For the analysis, we use the trading records of the aforementioned 10 listed companies of internet finance and regional banks from November 1, 2017 to October 12, 2018 for a total of 231 consecutive trading days. It is assumed that the observable information and unobservable information of the listed companies has been reflected in the changes of stock price, stock transaction amount and trading volume. For the daily trading data, price fluctuation is revealed from the close price, which reflects the change of one day return. The direct related factor of daily stock price is the trading volume and trading amount on that day. In addition, the amount of the circulation market value reflects the scale and value of the listed company, which is a comprehensive evaluation of the company's operating status by the market.

In model (\ref{eq2.3}), daily per stock return ($ShrRet$) (including dividend) is selected as the dependent variable. Daily per stock trading volume ($TrdVol$), daily per stock transaction amount ($TrdVal$) and daily per stock circulation market value ($MrkVal$) are selected as the explanatory variables. In order to increase the comparability between companies of different scales and market values, the three explanatory variables were standardized. Thus, the individual random effect panel data model without constant terms can be rewritten as:
\begin{equation}\label{eq26}
ShrRet_{it}  = \beta _1 TrdVal_{it}  + \beta _2 TrdVol_{it}  + \beta _3 MktVal_{it}  + {\bm{\lambda }}_i^{'} {\bf{f}}_t  + u_{it}.
\end{equation}

The proposed algorithm is used to implement the model (\ref{eq26}) together with the factor panel stochastic volatility model (\ref{eq2.2}) - (\ref{eq2.4}). According to number of factors choice the method proposed by Bai and Ng (2002), three common factors are selected. The estimates of the factor loadings and individual random effects are shown in table 3 and table 4 respectively.

From table 3, internet financial companies and traditional financial companies have obvious differences in estimated factor loadings. Among them, internet financial listed companies have generally higher loadings on the first common factor, while traditional financial listed companies have higher loadings on the second common factor. Since the common factors represent the common shocks or the common influencing factors on a group of stocks, it reveals that the influencing factors of these two classes of financial companies are completely different, or they are from different sources. On the other hand, traditional financial listed companies have both positive and negative factor loadings on the three factor loadings, indicating that their directions of impact are not consistent. However, for listed internet financial companies, the signs are all positive for the first factor loading and the second factor loading, and all negative for the third factor load, showing a consistent pattern. In conclusion, these two classes of financial listed companies show obvious difference behavioral from their transactions, which verifies that there is indeed a noticeable difference between their internal and external influential factors.

\begin{table}
\caption{Estimates and standard errors of panel data factor loadings of Internet financial and traditional financial companies.}
\begin{center}
\begin{tabular} {c|ccc||c|ccc}\hline\hline


Internet&	Fctor1&	Factor2&	Factor3&	Tradition&	Fctor1&	Factor2&	Factor3\\
                         Finance&&&&        Finance&&&\\

\hline
600570&	0.48295&	0.26396&	-0.39941&	601229&	0.24935&    -0.6055&    0.34556\\
300468&	0.48165&	0.2527&	   -0.33806&    601009&	-0.35094&	-0.80025&	0.10056\\
600446&	0.53242&	0.29306&	-0.43271&	002142&	-0.10866&	-0.61911&	0.21143\\
600588&	0.41928&	0.2797& 	-0.27535&	601997&	0.15192& 	-0.66763&	0.25843\\
002095&	0.40358&	0.28249&	-0.46229&	600919&	0.40977& 	-0.69211&	-0.18339\\
600599&	0.26668&	0.14423&	-0.16483&	600908&	0.45947&    -0.74334&	-0.20374\\
300300&	0.34265&	0.19878&	-0.29666&	600926&	0.31993&  	-0.73738&	0.13836\\
300295&	0.59382&	0.29622&	-0.28131&	002839&	0.35223&  	-0.62106&	-0.39002\\
002285&	0.44411&	0.23941&	-0.09751&	002807&	0.40254& 	-0.74668&	-0.34116\\
300178&	0.40544&	0.21892&	-0.31614&	601169&	-0.01292&	-0.69512&	0.16557\\


\hline\hline
\end{tabular}
\label{tab3}
\end{center}
\end{table}

While the factor loading analysis only reveals influence of unobservable factors, in the factor panel data stochastic fluctuation model, we can assess the influence of observable factors such as trading volume, transaction amount and current market value and their lag terms on the current stock return from the estimated regression coefficients. Using the Chinese stock market data, the estimates of model (\ref{eq26}) from the FFBS method are presented in table 4. Since all explanatory variables have been standardized, the coefficient of each explanatory variable is comparable for the two types of listed companies. As shown from table 4, the estimated coefficients of the stock circulation market value have relatively smaller standard errors, and all of them are significance at the significance level of 10$\%$. The estimated coefficients of these observable factors show no obvious difference in patterns for the two types of companies. Except for the bank of Shanghai (601229), there is a significant positive correlation between the current market value and the stock return. It shows that listed companies with larger current market values have a higher expected positive return during the observation period. The results do not seem to be consistent regarding the influence of trading volume and trading amount on the return for the two types of companies. However, except for the bank of Nanjing (601009), most stock trading volume and transaction amount are negatively correlated with the stock return. It shows that the market performance seems to differ between the high price stocks and the low price stocks.

\begin{table}
\caption{Estimates and standard errors of random coefficients for Internet financial and traditional financial companies.}
\begin{center}
\begin{tabular} {c|ccc||c|ccc}\hline\hline

Internet&	$TrdVol$&	$TrdVal$ &	$MrkVal$&	Tradition&	$TrdVol$&	$TrdVal$ &	$MrkVal$\\
                         Finance&&&&        Finance&&&\\

\hline

600570&	7.29E-03&	-0.00751&	0.030701&	601229&	-3.01E-02&	0.032915&	-0.00375\\
SE&	    1.86E-03&	0.001888&	0.000227&	SE&	    9.49E-03&	0.009507&	0.000948\\
300468&	-4.80E-03&	0.011402&	0.022815&	601009&	6.45E-05&	0.000153&	0.015969\\
SE&	     4.72E-03&	0.004782&	0.00155&    SE&	    2.84E-03&	0.00285&    0.00031\\
600446&	8.21E-03&	-0.00909&	0.031296&	002142&	5.09E-03&	-0.00478&	0.019807\\
SE&	    1.71E-03&	0.00176&    0.000439&	SE&	    2.18E-03&	0.002189&	0.000212\\
600588&	5.67E-03&	-0.0049&    0.036145&	601997&	3.23E-03&	-0.00342&	0.015774\\
SE&	    2.10E-03&	0.002107&	0.000429&	SE&	    1.72E-03&	0.001737&	0.000213\\
002095&	8.56E-04&	-0.00212&	0.030323&	600919&	-1.04E-02&	0.009408&	0.016956\\
SE&	    2.01E-03&	0.002051&	0.000403&	SE&	    1.20E-02&	0.012167&	0.001323\\
600599&	1.07E-03&	-0.00187&	0.021443&	600908&	1.09E-02&	-0.0124&    0.020783\\
SE&	    2.26E-03&	0.002284&	0.000378&	SE&	    1.66E-03&	0.001691&	0.000261\\
300300&	-2.52E-03&	0.005383&	0.020595&	600926&	2.38E-03&	-0.00234&	0.01458\\
SE&	    5.88E-03&	0.005877&	0.001016&	SE&	    6.91E-04&	0.000697&	0.000135\\
300295&	7.36E-03&	-0.00692&	0.028783&	002839&	-1.56E-02&	0.0213&	    0.003849\\
SE&	    2.14E-03&	0.002159&	0.000429&	SE&	    1.13E-02&	0.01144&    0.002175\\
002285&	1.13E-02&	-0.01066&	0.029251&	002807&	9.91E-03&	-0.00977&	0.02229\\
SE&	    2.87E-03&	0.002884&	0.000697&	SE&	    1.43E-03&	0.001442&	0.000291\\
300178&	-1.50E-03&	0.000133&	0.332127&	601169&	3.56E-03&	-0.00371&	0.010487\\
SE&	    3.29E-03&	0.000226&	0.005708&	SE&	    2.12E-03&	0.002139&	0.000208\\


\hline\hline
\end{tabular}
\label{tab4}
\end{center}
\end{table}

From the comparative analysis, Internet financial companies and traditional financial companies of concern are affected by various observable and unobservable factors. There seems to be similarity and dissimilarity with regard to the influential patterns with these factors. The effects of the observable factors seems to be similar for the market performance of the two types of companies, while the unobservable influence seems to different upon individual stocks.

Regarding the observable factors, the current market value of the company has a positive correlation with the stock return. Since the period from the end of 2017 to the end of 2018 selected here are stock price associated with a downward trend in China's stock market, the increase in trading volume or transaction amount maybe simply reflect the decline on the stock returns. From table 4, the estimated coefficients of individual random effects is negative, indicating a negative correlation between the return and the trading volume/trasaction amount. This demonstrates the potential purpose of using the factor panel data stochastic volatility model to predict the trend of company's stock return based on observable factors.

On the other hand, the influence of unobservable factors on the two types of listed companies seems to be noticeably different. Based on the estimates from factor decomposition for the unobservable factors, we may obtain practical implications of each common factor or a group of common factors and their associated influences. The results can be used to classify the company in order to help individual and institutional investors make reasonable investment decisions. Compared with other stochastic volatility models, the panel data factor stochastic volatility model containing both unobservable and observable factors provides an alternative tactic to analyze the investment value and market performance of a listed company.

\section{\sc {Conclusions and Further Research}}

In this paper, we proposed the panel data factor stochastic volatility model and discussed its estimation using the multivariate FFBS algorithm based on block sampling. Potential alternative methods for model implementation include the generalized moment method which has been widely used for the multivariate stochastic volatility model, quasi-maximum likelihood estimation, simulated maximum likelihood, important sampling technique, as well as other estimation methods which have been successfully applied to univariate and multivariate stochastic volatility models. Other filtering algorithms combining MCMC with Kalman filtering and particle filtering also can be considered. For high-frequency data, the reflection of jump, microstructure noise and realized volatility can also be considered when constructing estimation method.

For the empirical study, the Forward Filtering Backward Sampling algorithm is used to implement the new panel data model. The empirical study shows that the panel data factor stochastic volatility model, which considers the observable factors inside the market and the unobservable factors outside the market, can not only capture the impact of common shocks on the same type of stocks, but also differences between different types of stocks.  
Based on factor analysis in the literature shows that stock returns are main determined by macroeconomic factors, industrial factors and individual factors, we study the effects of observable factors including macroeconomic factor and industrial factors and unobservable factors representing individual factors. From the empirical analysis, we observe that the influence of observable factors and unobservable factors on stock returns shows different patterns. The observable factors seem to have similar effects on the two types of stocks from internet financial and traditional financial companies, while the unobservable individual factors seem to have different patterns of effects.

In addition to estimates of parameters and standard errors, future research can be done on model diagnosis for the panel data factor stochastic volatility model. Pitt and Shephard (1999) proposed four methods for the stochastic volatility model diagnostic, including the methods of logarithmic likelihood, standardized logarithmic likelihood, consistency residual, and distance measurement. Due to the large number of free parameters including latent factors included in the model, traditional methods such as AIC and BIC are not suitable for selection model estimated by MCMC method. For a complex model such as the factor model, the deviation information criterion (DIC) proposed by Spiegelhalter et al. (2002) can be difficult to computing. In future studies, the distance measurement method or an improved DIC can be considered. In addition, the model selection methods for spatial and temporal two-dimensional panel data models can be used for alternative Bayesian models.

\section*{Appendix:  }

\noindent {\bf The FFBS Algorithm for Panel Data Factor Stochastic Volatility Model}

The Forward Filtering and Backward Sampling (FFBS) algorithm was proposed by Carter \& Cohen (1994) and Fr\"{u}hwirth-Schnatter (1994) separately. Hore et al. (2010) documented the FFBS algorithm in a nonlinear state space model in detail. This paper mainly discusses the panel data factor stochastic volatility model estimation based on the multivariate FFBS algorithm.

For model (\ref{eq2.1}), the observable response and explanatory variables items (including the coefficients obtained by the joint estimate) can be merged and moved to the left hand, while the unobservable factors, the factor loadings and error components can be moved to the right hand of the model. After the re-parametrization and rearrangemeng, the model can be written as:
\begin{equation}\label{eqA.1}
{\bf{z}}_t  = {\bf{c}}_t  + {\bf{b}}_i {\bf{\tilde h}}_t  + {\bf{e}}_t.
\end{equation}
Let ${\bf{h}}_t  = {\bf{b}}_i {\bf{\tilde h}}_t$, and assume that ${\bf{h}}_{t - 1}^*  = ({\bf{I}},{\bf{h}}_{t - 1} )$. Since the unobservable factors ${\bf{b}}_i$ are uncorrelated with the time variable, we can obtain model (\ref{eq22}) and (\ref{eq23}). To predict the implicit volatility block, in model (\ref{eq24}), neglecting the known parameters and data information gives the conditional probability density:
\begin{equation}\label{eqA.2}
\pi ({\bf{h}}|{\bf{I}}_T ) = p({\bf{h}}_T |{\bf{I}}_T )\prod\limits_{t = 1}^T {p({\bf{h}}_t |{\bf{h}}_{t + 1} ,\mathscr{I}_t )},
\end{equation}
with the Information set $\mathscr{I}_t$ and ${\bf{I}}_T$ defined earlier.

Since
\begin{equation}\label{eqA.3}
p({\bf{h}}_{T - 1} ,{\bf{h}}_T |{\bf{I}}_T ) = p({\bf{h}}_T |{\bf{I}}_T )p({\bf{h}}_{T - 1} |{\bf{h}}_T ,{\bf{I}}_T ),
\end{equation}
the backward sampling process of (\ref{eqA.2}) can be written as:
\begin{equation}\label{eqA.4}
p({\bf{h}}_1 ,{\bf{h}}_2 , \cdots ,{\bf{h}}_T |{\bf{I}}_T ) = p({\bf{h}}_{T - 1} ,{\bf{h}}_T |{\bf{I}}_T )\prod\limits_{t = 1}^{T - 2} {p({\bf{h}}_t |{\bf{h}}_{t + 1} ,\mathscr{I}_t )}.
\end{equation}
The estimation of high-dimensional implicit volatility terms ${\bf{h}}_t$ can be done in two steps: forward filtering and backward sampling.

\subsubsection{Forward Filtering}
To use the information collected at timing $t$ to predict the state of future timing, we can add the state variables ${\bf{h}}_{t + 1}$ at timing $t+1$ to the state equation (\ref{eqA.3}), and obtain the full conditional density of joint states as:
\begin{equation}\label{eqA.5}
p({\bf{h}}_{t - 1} ,{\bf{h}}_t ,{\bf{h}}_{t + 1} |\mathscr{I}_t ) = p({\bf{h}}_{t - 1} ,{\bf{h}}_t |\mathscr{I}_t )p({\bf{h}}_{t + 1} |{\bf{h}}_{t - 1} ,{\bf{h}}_t ,\mathscr{I}_t ).
\end{equation}
The $\mathscr{I}_t$ includes the full information of model parameters and data ${\bf{I}}_T$. From the property of Markov chains, we have
\begin{equation}\label{eqA.6}
p({\bf{h}}_{t + 1} |{\bf{h}}_{t - 1} ,{\bf{h}}_t ,\mathscr{I}_t ) = p({\bf{h}}_{t + 1} |{\bf{h}}_t ,\mathscr{I}_t ) = p({\bf{h}}_{t + 1} |{\bf{h}}_t ).
\end{equation}
The first term at the RHS of equation (\ref{eqA.5}) can be written as
\begin{equation}\label{eqA.7}
p({\bf{h}}_{t - 1} ,{\bf{h}}_t |\mathscr{I}_t ) = p({\bf{h}}_{t - 1} |{\bf{h}}_t ,\mathscr{I}_t )p({\bf{h}}_t |\mathscr{I}_t ),
\end{equation}
where
\begin{equation}\label{eqA.8}
\begin{aligned}
p({\bf{h}}_{t - 1} |{\bf{h}}_t ,\mathscr{I}_t ) &= p({\bf{h}}_{t - 1} |{\bf{h}}_t ,\mathscr{I}_{t - 1} ,{\bf{z}}_t ) \\
&= p({\bf{h}}_{t - 1} |{\bf{h}}_t ,\mathscr{I}_{t - 1} ).
\end{aligned}
\end{equation}
The last step of (\ref{eqA.8}) is true since ${\bf{h}}_{t - 1}$ and ${\bf{z}}_t$ are conditionally independent. Combining (\ref{eqA.5})-(\ref{eqA.8}), we have
\begin{equation}\label{eqA.9}
p({\bf{h}}_t ,{\bf{h}}_{t + 1} |\mathscr{I}_t ) = p({\bf{h}}_t |\mathscr{I}_t )p({\bf{h}}_{t + 1} |{\bf{h}}_t ).
\end{equation}
After introducing the information of parameters and data at timing $t+1$, (\ref{eqA.9}) evolved into a renewal process. At this point, the forward filtering algorithm of state equation (\ref{eq23}) is realized by transfer rule (\ref{eq24}), where
\begin{equation}\label{eqA.10}
({\bf{h}}_t |{\bf{h}}_{t + 1} ,{\bm{\Sigma }}_e ,{\bf{\Sigma }}_\nu  ,\mathscr{I}_t ) \sim N(({\bm{\alpha }}^{'} {\bm{\Sigma }}_\nu ^{ - 1} {\bm{\alpha }} + {\bf{m}}_t^{ - 1} )^{ - 1} ({\bm{\alpha }}^{'} {\bm{\Sigma }}_\nu ^{ - 1} {\bf{h}}_{t + 1}  + {\bf{D}}_t ),({\bm{\alpha }}^{'} {\bm{\Sigma }}_\nu ^{ - 1} {\bm{\alpha }} + {\bf{m}}_t^{ - 1} )^{ - 1} ).
\end{equation}
Moreover, the Kalman filtering algorithm of the nonlinear state equation is given by
\begin{equation}\label{eqA.11}
\begin{aligned}
({\bf{h}}_{t - 1} ,{\bf{h}}_t |\mathscr{I}_{t - 1} ) &\sim N({\bf{m}}_{t|t - 1} ,{\bf{D}}_{t|t - 1} )
\\
{\bf{m}}_{t|t - 1} & = \left( {{\bf{h}}_{t - 1|t - 1} ,{\bf{h}}_{t|t - 1} } \right)
\\
{\bf{D}}_{t|t - 1} & = \left( {\begin{array}{*{20}c}
   \begin{array}{l}
 {\bf{\Sigma }}_{t - 1|t - 1}  \\
 {\bm{\alpha \Sigma }}_{t - 1|t - 1}  \\
 \end{array} & \begin{array}{l}
 {\bm{\alpha \Sigma }}_{t - 1|t - 1}  \\
 {\bf{\Sigma }}_{t|t - 1}  \\
 \end{array}  \\
\end{array}} \right),
\end{aligned}
\end{equation}
where $\bullet | \bullet$ represents the state transition equations, the corresponding ${\bf{m}}_{t|t - 1}$ and ${\bf{D}}_{t|t - 1}$ terms are conditional expectation and conditional variance-covariance matrices. Filtered by equation (\ref{eqA.3}), the posterior samples of the implied volatility ${\bf{h}}_t$ can be obtain in sequence.

\subsubsection{Backward Sampling}

The backward sampling process of equation (\ref{eqA.4}) is designed to smooth the state variables. From the property of Markov chains, we have
\begin{equation}\label{eqA.12}
\begin{aligned}
p({\bf{h}}_t |{\bf{h}}_{t + 1} , \cdots ,{\bf{h}}_T ,\mathscr{I}_T ) &= p({\bf{h}}_t |{\bf{h}}_{t + 1} ,{\bf{I}}_{T + 1} ,\mathscr{I}_T )\\
& = p({\bf{h}}_t |{\bf{h}}_{t + 1} ,\mathscr{I}_{T + 1} ).
\end{aligned}
\end{equation}
On the other hand, we can obtain
\begin{equation}\label{eqA.13}
p({\bf{h}}_{t - 1} ,{\bf{h}}_t ,{\bf{z}}_t |{\bf{h}}_{t + 1} ,{\bf{z}}_{t + 1} ,\mathscr{I}_{t + 1} ) = p({\bf{h}}_{t - 1} ,{\bf{h}}_t ,{\bf{z}}_t |{\bf{h}}_{t + 1} ,{\bf{z}}_{t + 1} ).
\end{equation}

In order to realize the posterior joint sampling, assuming that the joint distribution of $({\bf{h}}_{t - 1} ,{\bf{h}}_t )$ is the multivariate normal distribution, the block samples of the implied volatility terms can be extracted from the distribution of (\ref{eqA.11}).

\begin{flushright}
\qed
\end{flushright}

\noindent
{\bf Compliance with Ethical Standards:} This article does not contain any study on human subjects or animals.
\bigskip

\noindent
{\bf Funding:} This research was funded by National Natural Science Foundation of China ( 714711730 , 71873137 , 71271210); Fang's study was funded by The Philosophy and Social Science Fund of Anhui (AHSKY2015D53).

\bigskip
\noindent
{\bf Disclosure of potential conflicts of interest:} None


\end{document}